\documentclass[12pt]{interact}

\usepackage{nameref}

\usepackage{mathptmx} % Times New Roman font
\usepackage{graphicx}

\newcommand\wordcount{
    \immediate\write18{texcount -sum -1 \jobname.tex > \jobname-words.sum }
    \input{\jobname-words.sum} words
}

\usepackage{natbib}
\usepackage{caption}
\usepackage{xcolor}
\usepackage{enumerate}
\usepackage{epstopdf}% To incorporate .eps illustrations using PDFLaTeX, etc.
\usepackage[caption=false]{subfig}% Support for small, `sub' figures and tables
\usepackage{algpseudocode}
\bibpunct[, ]{(}{)}{;}{a}{,}{,}% Citation support using natbib.sty
\usepackage{lineno}
\usepackage{amsmath}

\usepackage{algorithm}
\usepackage{algpseudocode}
\usepackage{array}
\usepackage{url}
\usepackage{soul}

\theoremstyle{plain}% Theorem-like structures provided by amsthm.sty

\theoremstyle{definition}

\theoremstyle{remark}

\usepackage{geometry}
\setcitestyle{authoryear,open={(},close={)}}

\begin{document}

\title{Identifying rich clubs in spatiotemporal interaction networks}

% ORCID
% Jacob Kruse: 0000-0001-7752-3190
% Song Gao: 0000-0003-4359-6302

\author{\name{Jacob Kruse \textsuperscript{a}, Song Gao \textsuperscript{a,*}, Yuhan Ji \textsuperscript{a}, Keith Levin \textsuperscript{b}, Qunying Huang \textsuperscript{a}, Kenneth R. Mayer \textsuperscript{c} } 
\affil{\textsuperscript{a} Department of Geography, University of Wisconsin-Madison, USA \\ \textsuperscript{b} Department of Statistics, University of Wisconsin-Madison, USA \\
\textsuperscript{c} Department of Political Science, University of Wisconsin-Madison, USA}
}

\thanks{*Corresponding author: song.gao@wisc.edu}
\maketitle

\begin{abstract}
Spatial networks are widely used in various fields to represent and analyze interactions or relationships between locations or spatially distributed entities. While existing studies have proposed methods for hub identification and community detection in spatial networks, relatively few have focused on quantifying the strength or density of connections shared within a community of hubs across space and time. Borrowing from network science, there is a relevant concept known as the `rich club' phenomenon, which describes the tendency of `rich' nodes to form densely interconnected sub-networks. Although there are established methods to quantify topological, weighted, and temporal rich clubs individually, there is limited research on measuring the rich club effect in spatially-weighted temporal networks, which could be particularly useful for studying dynamic spatial interaction networks. To address this gap, we introduce the spatially-weighted temporal rich club (WTRC), a metric that quantifies the strength and consistency of connections between rich nodes in a spatiotemporal network. Additionally, we present a unified rich club framework that distinguishes the WTRC effect from other rich club effects, providing a way to measure topological, weighted, and temporal rich club effects together. Through two case studies of human mobility networks at different spatial scales, we demonstrate how the WTRC is able to identify significant weighted temporal rich club effects, whereas the unweighted equivalent in the same network either fails to detect a rich club effect or inaccurately estimates its significance. In each case study, we explore the spatial layout and temporal variations revealed by the WTRC analysis, showcasing its particular value in studying spatiotemporal interaction networks. This research offers new insights into the study of spatiotemporal networks, with critical implications for applications such as transportation, redistricting, and epidemiology.

\end{abstract}


\newpage

\section{Introduction}\label{sec:introduction}

Spatial networks are widely used across various fields, such as in geography, transportation, public health, environment and ecology, to represent and analyze connections and flows between locations or spatially distributed objects or entities \citep{barthelemy2011spatial,gersmehl1970spatial,bian2024introduction}. In such networks, the \textit{edge weights} encode the volume of flows or interactions between \textit{nodes} (representing locations, spaces, and places), reflecting the effects of geographic distance, adjacency, and other processes on the strength or likelihood of connections between two locations or geographical entities \citep{roy2004spatial,zhong2014detecting,andris2016integrating,clarke2018spatial,xu2022beyond}. 
One common use for spatial interaction networks is to model human mobility flows, and research across various disciplines has developed and utilized network methods to analyze human mobility flows in diverse ways. For instance, community detection methods have been applied to identify spatially connected geographic communities within these networks based on interaction strengths \citep{gao2013discovering,chen2015finding,liang2022region2vec,wang2021network}. Recent studies have extended these approaches to detect dynamic communities in spatiotemporal interaction networks \citep{kang2022measuring,jia2022dynamical,zhao2023dynamic}, revealing patterns in human mobility dynamics over time.

In another line of research, patterns in human mobility flows and their changes in response to disruptions such as COVID-19 have been analyzed to understand spatiotemporal phenomena like disease transmission. For example, \cite{chin2024networked} examined how urban mobility dynamics in one city evolved during the COVID-19 pandemic, while \cite{kang2020multiscale} and \cite{hou2021intracounty} analyzed how human mobility flows changed in response to COVID-19 infection at different geographic scales and over time. Since human mobility flow datasets are typically aggregated to geographic units; the nodes in such spatial interaction networks represent statistical areas rather than distinct entities directly corresponding to cities, towns, or other geographic features. The scale and specific boundary locations of these aggregated regions can significantly influence subsequent analyses—a challenge known as the modifiable areal unit problem (MAUP) \citep{openshaw1984modifiable,atkinson2000spatial}.
One common approach to addressing the MAUP problem is to perform analyses at multiple spatial scales, which helps to reveal how patterns change across different levels of aggregation and reflect the varying spatial processes at work \citep{oshan2022scoping,quattrochi2023scale}. Given the additional complexity of analyzing spatial interaction patterns across scales, such studies often represent human mobility flows as undirected spatial networks. 

A relatively new application of human mobility network is in redistricting tasks. Redistricting is a type of regionalization task that involves dividing a larger geographic area into smaller discrete regions, often with specific goals such as equalizing population or ensuring fair representation. Recent studies have utilized human mobility networks to produce and evaluate district plans based on how well they align with underlying human mobility patterns across geographic communities. For example, \cite{liang2022region2vec,liang2025geoai} and \cite{wang2021network} used human mobility networks to design health service areas that account for ease of access and existing mobility patterns. In the context of political redistricting, \cite{kruse2023bringing} evaluated district plans based on spatial interaction networks, with plans having relatively more intra-district flows considered superior to those having higher inter-district flows.
One aspect of political redistricting research that remains under-explored is the use of dynamic human mobility flows to understand district cores. The preservation of district cores is a frequently cited but ambiguously defined criterion in redistricting research \citep{eckman2021congressional, yablon2022gerrylaundering}. A quantitative framework based on dynamic human movements could provide valuable insights into defining and understanding these core regions as spatial interaction communities, which can also contribute to the geographic process-oriented regionalization based on spatiotemporal data~\citep{zhang2024advancing}.

To analyze district cores using human mobility flows, it is necessary to employ a method capable of identifying core regions within the dynamic human mobility network (i.e., spatial connections or interaction strengths change over time). One promising concept from network science, known as the `rich club' phenomenon~\citep{Colizza2006}, could be particularly useful for identifying district cores in spatial interaction networks. Unlike community detection, which partition the entire (spatial) network to cohesive sub-networks~\citep{newman2004finding}, the rich club phenomenon describes a network property that quantifies the tendency of `rich' nodes—those with high connectivity or influence—to form densely interconnected sub-networks. In this context, rich club analysis can serve as a method in geographic networks to quantify and evaluate the strength of connections or flows among a subset of influential nodes within the networked geographic phenomena such as population migration~\citep{bian2024introduction,koylu2022measuring}.
Various definitions of the rich club exist \citep{zhou2004rich,Colizza2006,alstott2014unifying,pedreschi2022temporal}, with the topological rich club (TopoRC) and weighted rich club (WRC) being two of the most commonly studied types. For instance, in scientific collaboration networks, the TopoRC captures how influential researchers are more likely to collaborate with each other than with less influential peers \citep{Colizza2006}. In contrast, the WRC quantifies the weight or strength of connections between rich nodes, such as in global airline networks, where major airports exchange a higher volume of flights compared to smaller airports \citep{alstott2014unifying}.
Considering that many human mobility networks exhibit temporal variation \citep{kang2020multiscale}, it is important to explore the temporal stability of the rich club effect when studying human mobility networks.
To identify temporally consistent rich club behavior in time-varying networks, \cite{pedreschi2022temporal} introduced the temporal rich club (TRC) coefficient. However, this method relies solely on temporal edge connections rather than edge weights, and so it might also be referred to as the topological TRC. Since the edge weights are not incorporated, this method might be unsuitable for studying weighted spatial interaction networks based on dynamic human mobility flows.

To address this gap, this paper introduces a novel method for measuring the rich club effect in undirected, weighted, and dynamic spatiotemporal networks. The proposed rich club methodology is then applied a time-varying human mobility flow network to identify congressional district cores. Additionally, building on existing works that analyze changes in human mobility flows in response to COVID-19, this study explores the rich club effect in dynamic mobility networks before and after the onset of COVID-19. This analysis reveals the extent to which human mobility flows were concentrated within important sub-networks during the pandemic—an understudied element of COVID-19 human mobility research.
To understand how the MAUP affects rich club analyses, we perform the rich club analyses on spatiotemporal networks at different geographic scales. It is worth noting that while we focus on the study of absolute space in this study, the proposed WTRC can also be applied to identify rich clubs in relational space (such as environmental governance network). Many interesting discussions on such distinctions and the diversity of geographic networks can be found in the Special Issue: `Networks' of Annals of the American Association of Geographers~\citep{bian2024introduction}.

The remainder of the paper is organized as follows. In the Related Work section, we review the relevant definitions and metrics for identifying rich clubs, along with the randomization methods used to differentiate between topological, weighted, and temporal rich club effects. Next, in Methods section, we introduce the new methodology for weighted temporal rich club (WTRC) quantification and normalization. The Experiments and Results section presents two experiments conducted on different dynamic spatial interaction networks to demonstrate the utility of the proposed WTRC method at different geographic scales. In the Discussion section, we discuss some of the limitations and implications of this work. Finally, the Conclusion and Future Work section provides concluding remarks and suggests directions for future research.

\section{Related Work} \label{sec:relatedwork}
\subsection{Rich club analysis in spatial interaction networks}

As described in the introduction, the description of spatial networks~\citep{barthelemy2011spatial} aligns with the definition of geographic networks provided by \cite{bian2024introduction}, where locations serve as nodes, edges represent connections or interactions between these locations, and spatial autocorrelation is an underlying characteristic of such networks.
Following both of these definitions, spatial interaction networks can be called spatially-embedded or spatially-weighted networks, reflecting the fact that the nodes are embedded in geographic space and that the edges are weighted by flows or interactions over space. We note, however, that this is different from a similar use of the term `spatial weights' (another common term in the geography literature), which is often used in the measures of spatial autocorrelation and in the spatial regression models, where the weight represents the strength of correlation between two points at a given spatial distance \citep{getis2007reflections,getis2009spatial}. Even though the rich club literature has so far not explicitly employed the term spatial interaction networks, several rich club studies use maps with the explicit purpose of further understanding the spatial interactions of the rich clubs identified in their analyses. For example, \cite{pedreschi2022temporal} map the rich club of U.S. airports, demonstrating via spatial proximity that many of the rich club members are, in fact, reliever airports to larger hubs. Similarly, \cite{zhang2021unveiling} employ various maps to understand the spatial distribution of passenger flows across different times and for various rich club members (i.e., transportation stations). As these networks are indeed spatial interaction networks, rich club analysis can reveal that the discovered hubs do indeed exhibit spatial autocorrelation, where spatially-influenced factors such as travel time affect the extent to which rich nodes direct flows to each other.

Having established the suitability of analyzing spatial interaction networks using rich club methodology, we now turn to a review of the relevant rich club literature from network science.

\subsection{Static rich clubs} 
\subsubsection{Topological rich club}
\label{sec:topo_rc}

The most basic type of rich club analysis can be performed in static, unweighted networks, such as a network of airports connected by direct flight routes \citep{Colizza2006}. To quantify the extent of connectivity among highly connected nodes in such networks, \cite{Zhou2004} introduced the topological rich club (TopoRC) coefficient, denoted as $\phi$. The TopoRC coefficient is defined as the ratio of the actual number of connections between rich nodes to the maximum number of possible connections among them. Rich nodes are typically identified as those with a degree greater than or equal to a threshold $k$. With $N$ representing the number of rich nodes and $E$ representing the number of edges shared among them, the rich club coefficient is calculated using the following formula:

\begin{equation}
\phi = \frac{2E}{N(N-1)}
\label{eq:1}
\end{equation}

Since each network has a range of richness values that form a \textit{richness sequence} (e.g., the set $k_{1}, k_{2}, \dots, k_{n}$ for all nodes in the network), the rich club coefficient can be calculated at different \textit{richness thresholds} or \textit{levels}. A richness threshold is the specific richness value used to determine whether a node qualifies as a rich club member—if a node's richness value is greater than or equal to the threshold, the node is considered a rich club member.
A high rich club coefficient indicates the presence of an `oligarchy' in a network, where a few influential members disproportionately connect with each other,
though this is complicated by the fact that even random graphs can have rich club effects due to edge placement.
To accurately quantify the organizing structures in a real network, the structural correlations (finite-size effects) inherent in a random graph must be discounted from the rich club coefficient. 
This is done by normalizing the rich club coefficient against that of a comparable random graph having the same richness sequence, with the division allowing for fraction reduction:

\begin{equation}
\label{eq:topo_norm}
\phi_{norm} = \frac{\phi}{\phi_{rand}} = \frac{E}{E_{rand}}
\end{equation}

\begin{figure}[]
    \centering
    \includegraphics[width=1\textwidth]{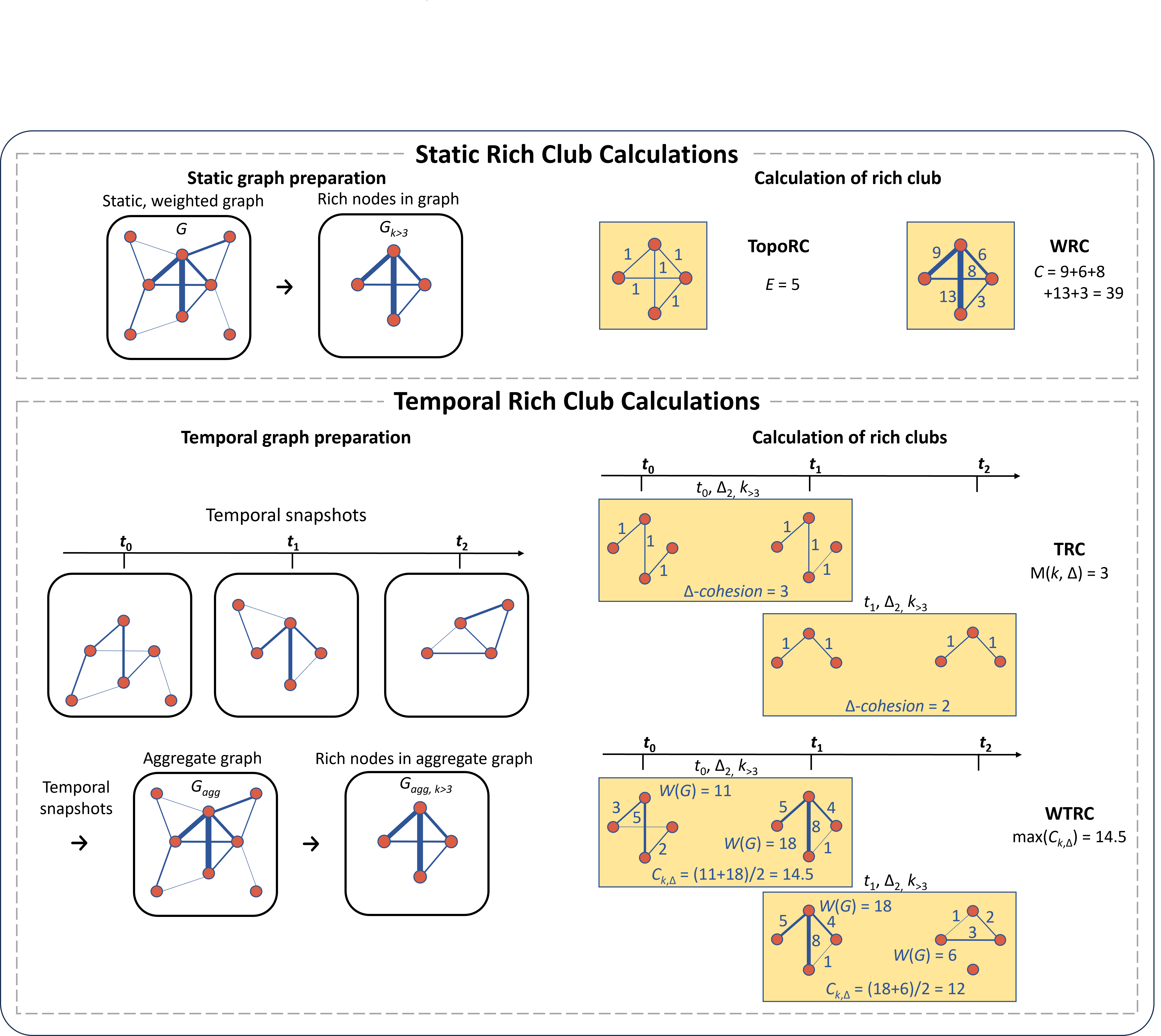}
    \caption{Calculations for all rich club types. For simplicity, only the numerator portion of the rich club calculation--derived from the graph being analyzed--is shown. The denominator portion follows the same calculation as the numerator but applied to a randomized version of the graph.}
  \label{fig:all_rich_clubs}
\end{figure}

\begin{figure}[]
    \centering
    \includegraphics[width=1\textwidth]{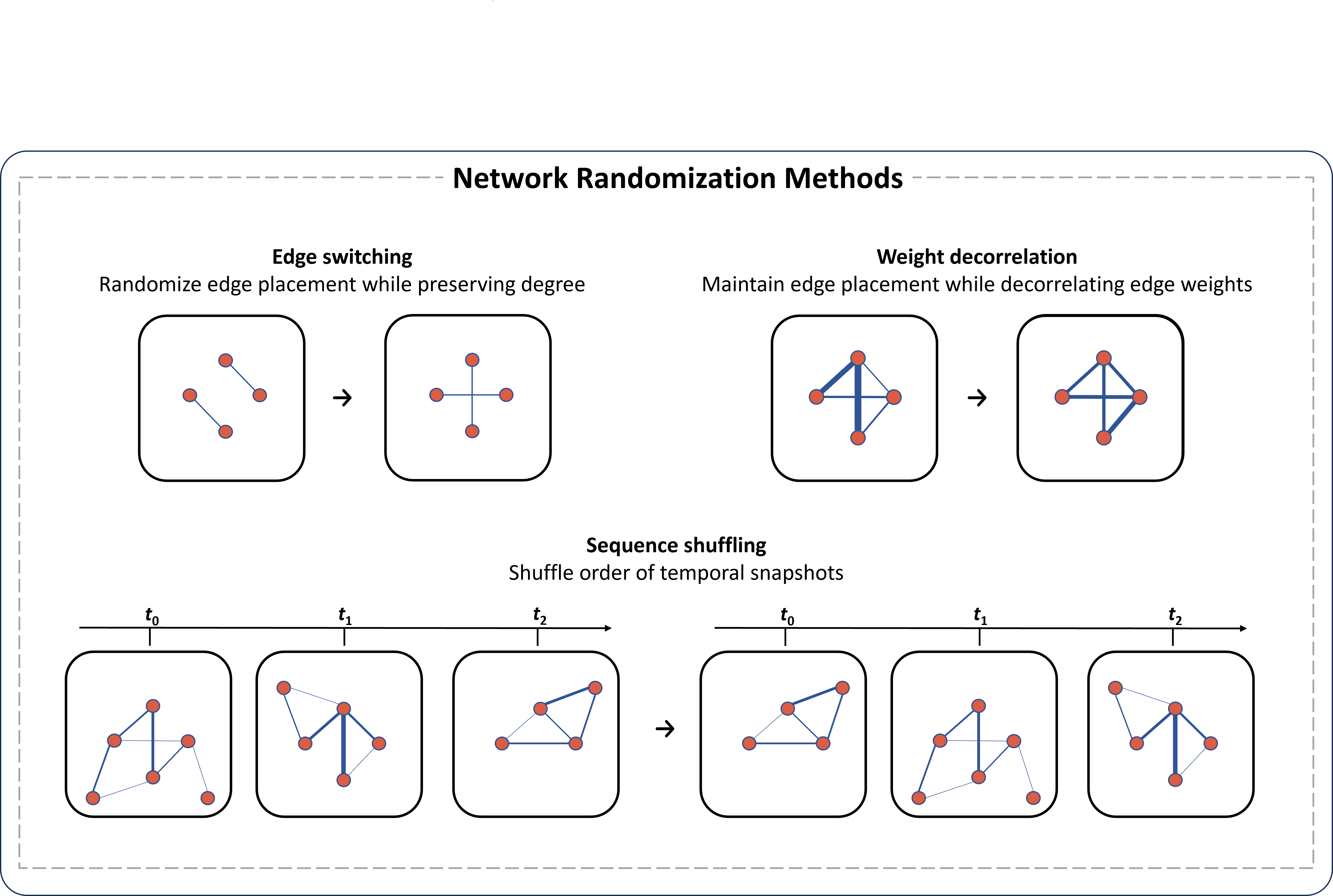}
    \caption{Methods for network randomization. Edge switching changes the placement of edges in the graph while maintaining every node's degree. Weight decorrelation randomizes weight allocation while maintaining graph topology. Sequence shuffling randomly shuffles the order of temporal snapshots in a temporal network, while maintaining the weight allocation and topology within each snapshot.} 
  \label{fig:randomization_methods}
\end{figure}

Where $\phi_{\text{rand}}$ is the rich club coefficient calculated on the random graph, and $E_{\text{rand}}$ is the edge count between rich nodes in the same graph. An example calculation of the TopoRC can be seen in Figure \ref{fig:all_rich_clubs}.
Values greater than 1 for $\phi_{\text{rand}}$ indicate the presence of a significant rich club, while values less than 1 indicate that there is less of rich club effect in the graph of interest than what would be expected in random graph.
The random graph, or null model, used for normalizing the TopoRC is typically generated using a variant of the edge-switching algorithm \citep{Colizza2006,alstott2014unifying}. This Monte Carlo method preserves the degree sequence while randomizing edge placement \citep{milo2003uniform,mcauley2007rich,roberts2000simple}, as illustrated in Figure \ref{fig:randomization_methods}. By maintaining the degree distribution, this method generates null models that can capture the range of degree distributions commonly found in both spatial and non-spatial networks \citep{gastner2006spatial,barthelemy2003crossover}.

\subsubsection{Weighted rich club}
\label{sec:wrc_calc}

A natural extension of the TopoRC would be a version of the rich club for edge-weighted networks, and indeed several works have proposed WRC definitions and applied them to spatial networks, including air transit passenger flows \citep{opsahl2008prominence, alstott2014unifying} and trade volumes between countries \cite{zlatic2009rich}.
However, as \cite{alstott2014unifying} describe, the use of normalization allows all the various WRC definitions to be expressed with the generalized form:
\begin{equation}
\label{eq:weighted_simplified}
\phi_{norm} = \frac{\phi}{\phi_{rand}} = \frac{C}{C_{rand}}
\end{equation}

Where the weighted connectedness of the rich club members, $C$, is defined as the sum of edge weights for edges that go between two rich club nodes (see Figure \ref{fig:all_rich_clubs} for an example). $C_{rand}$ is the same measurement calculated on an appropriate null model. Crucially, this generalized equation for the WRC requires that $C_{rand}$ have the same maximum weighted connectedness as $C$,
implying that appropriate constraints must be applied to the randomization of topological and weighted aspects of the graph during the production of the null model.
Since the WRC is based on weights assigned to edges, there is an additional challenge of distinguishing the effects of edge placement from the effects of weight allocation. As \cite{alstott2014unifying} detail, proper normalization can indeed distinguish between topological, weighted, and mixed rich club effects.

To quantify the TopoRC effect in a weighted network, the edge weights in the weighted network are all set to one, and then the normalized TopoRC coefficient is measured as described in the Topological rich club section, a process which includes generation of the null model with the edge switching algorithm.
To detect the WRC effect, specifically, \cite{alstott2014unifying} employ a weight decorrelation method based on randomized controls with the same topology but shuffled weights. 
This method, originally described by \cite{serrano2006correlations}, maintains network topology while producing a maximally random network with respect to the weights (see Figure \ref{fig:randomization_methods}).
To measure the mixed rich club effect in a weighted network, both edge placement and weight allocation are randomized using edge switching and weight decorrelation, and the mixed weighted and topological rich club effect is measured on both the real network and the randomized network.

\subsection{Temporal rich club phenomenon}
\label{sec:ttrc}
Another logical extension of the rich club is to characterize its dynamic nature over time. In one earlier work on this topic, \cite{zhang2021unveiling} compared outgoing-strength and incoming-strength global and local rich club coefficients for passenger railway transit systems for different hours throughout the day, with the related networks being directed, weighted, and temporal. However, this work did not consider connections, between the same set of rich nodes, that were consistent for the entire day. Rather, rich nodes were selected for each snapshot, independently. 
In a different work, \cite{pedreschi2022temporal} proposed the TRC as a way to quantify the rich club effect in undirected, unweighted dynamic networks. The TRC quantifies the degree to which rich clubs are tightly and simultaneously connected for some duration of time in a dynamic network. Using a variety of datasets, such as the U.S. air transportation network and regions of the brain, this work shows how the measurement of temporal rich clubs can reveal important sub-networks that are otherwise obscured in a static rich club framework. 

As the weighted temporal rich club definition that we will propose later is based on the TRC, we here describe the TRC graph construction and algorithm more explicitly. 
Consider a temporal, unweighted network composed of instantaneous snapshots on the interval $[1,T]$,
% of the network at different timestamps on the interval $[1,T]$, 
where a temporal edge at time $t$ between nodes $i$ and $j$ is represented as an edge in the instantaneous graph $G_{t}$. 
From these snapshots, a static, time-aggregated graph $G(N,E)$ can be composed, where the edge weight $w_{ij}$ is the sum of temporal edges between nodes $i$ and $j$ over the interval $[1,T]$. Accordingly, node strength $s_{i}$ is the total number of temporal edges associated with node $i$ over the same interval, and node degree $k_{i}$ in $G$ is the count of distinct nodes that node $i$ has interacted with at least once on the interval $[1,T]$.
Using the temporal network defined by both $G$ and the associated temporal snapshots, the algorithm for calculating the TRC coefficient works in the following way. The number of edges between rich nodes that are stable over $[t,t+\Delta]$ is normalized by the maximal connection of the same nodes over $[t,t+\Delta]$, where duration $\Delta$ is the number of temporal snapshots to be considered. This yields a $\Delta-cohesion$ value for a given $t$, richness threshold $k$, and duration $\Delta$: 

\begin{equation}
\label{eq:delta_coh}
    \Delta-cohesion = \epsilon_{>k}(t, \Delta)
\end{equation}

Where the richness threshold $k$ is a property of the time-aggregated graph $G$, rather than a property of a particular snapshot, and $\epsilon_{>k}(t, \Delta)$ is total number of edges shared between nodes with aggregated richness (e.g., time-aggregated degree) larger than $k$ that are stably present for duration $\Delta$ starting at time $t$. 

To scan the entire temporal network for significant patterns, \cite{pedreschi2022temporal} use the $\Delta-cohesion$ value in moving window fashion for every time $t$ on the interval $[1,T-\Delta]$ (see Figure \ref{fig:all_rich_clubs} for an example).
% For a given richness threshold $k$, the max $\Delta$-cohesion value found, $M(k,\Delta)$, across $[1,T-\Delta]$ is reported. 
As a function of $\Delta-cohesion$, the TRC coefficient is then defined as the maximal density of temporal edges observed for duration $\Delta$ found on the interval $[1,T-\Delta]$ among nodes of aggregated richness larger than $k$~\citep{pedreschi2022temporal}:

\begin{equation}
M(k, \Delta) = \max_t \epsilon_{>k}(t, \Delta)
\label{eq:max_delta_coh}
\end{equation}

The TRC coefficient $M(k, \Delta)$ can then be calculated for multiple richness ($k$) thresholds and $\Delta$ durations, allowing for the production of a two-dimensional results array with the maximal density of temporally-stable edges found with across different richness thresholds and lengths of time. Finally, as with the TopoRC and WRC, $M(k, \Delta)$ should be normalized, yielding:

\begin{equation}
M(k, \Delta)_{norm} = \frac{M(k, \Delta)}{M(k, \Delta)_{rand}}
\label{eq:normalized_M_trc}
\end{equation}

Where $M(k, \Delta)$ is the TRC coefficient in the real network, and $M(k, \Delta)_{rand}$ is the TRC coefficient in the randomized network.
Though $M(k, \Delta){rand}$ is essential for normalization, the most appropriate randomization strategy for generating the null model from which $M(k, \Delta)_{rand}$ is calculated remains under-explored.

As TRC definition relies on both temporal and topological aspects of the network, and appropriate randomization procedure will randomize both temporal and topological aspects of the network in the production of the null model. Indeed, the existing works on the TRC \citep{pedreschi2022temporal,li2023temporal} both use a randomization procedure that shuffles the simultaneity and stability of interactions, while preserving the number of temporal edges at each time step and the total number of temporal edges between each pair of nodes. 
In this model, referred to as timestamp shuffling \citep{gauvin2022randomized}, 
the randomization is achieved by swapping pairs of events (i.e., temporal edges) between edge timelines while conserving their timestamp sequences. Thus, the total number of temporal edges is conserved for each node pair, as is cumulative activity, i.e., the total number of temporal edges in the snapshot graphs from $t = 1$ through $t=T$. While timestamp shuffling does randomize the simultaneity and stability of connections (i.e., the temporal aspect), it is less clear that it sufficiently randomizes the topological aspect. From the topological perspective, each snapshot graph is \textit{not} randomized by shuffling link placement while maintaining node degree, as is done in edge switching.
Rather, the timestamp shuffling method merely swaps temporal edges between node pairs, without sampling uniformly from all possible node pair placements. Although this preserves the total number of temporal links for a given node pair over $[1,T]$, it does not necessarily maintain the degree sequence in each of the snapshots, such that the randomized snapshots no longer have the same level of finite-size effects as the original snapshots.

One potential alternative to timestamp shuffling could be a combination of sequence shuffling, in which the temporal order of the snapshot graphs is shuffled at random \citep{gauvin2022randomized}, and edge switching within each snapshot (see Figure \ref{fig:randomization_methods}). This combination could provide a randomized temporal network that will 1) have broken up significant temporal patterns by shuffling the order of the snapshots, thereby establishing a baseline for rich edge stability and simultaneity over duration $\Delta$, and 2) have randomized the topology sufficiently to provide a baseline for the level of structural correlation present in each individual snapshot. With this combined method of randomization, the normalized TRC effect would be a reflection of temporally stable topological rich clubs, rather than rich clubs that are stable temporally but not necessarily more structurally interconnected at the snapshot level than expected by chance.

\subsection{Significance testing}
\label{sec:temp_ran}
Whether it be for the topological, weighted, or temporal rich club, the normalization process is essentially dividing the rich club coefficient for the real network by the rich club coefficient for the null model. A natural extension of this process is the application of hypothesis testing, where the observed rich club coefficient is compared to a sample distribution of null models, as proposed by \cite{jiang2008statistical}. Accordingly, a two-tailed test or similar can be used to test for statistical significance \citep{witte2017statistics}. With a suitable number of randomized networks for hypothesis testing, Equation \ref{eq:topo_norm} can be modified to:

\begin{equation}
\label{eq:topo_norm_ave}
\phi_{norm} = \frac{\phi}{\langle \phi_{rand} \rangle}
\end{equation}

Where the denominator represents the average rich club coefficient derived from a set of null models. For the rest of this paper, any normalized rich club coefficient values presented should be understood to be the results of multiple null models.

\section{Methods}\label{sec:methods}
To study the rich club effect in undirected, weighted, and time-varying spatial interaction networks—such as those based on dynamic human mobility flows over time—a weighted temporal rich club definition is essential. As no existing method in the literature quantifies the temporal rich club effect in weighted networks, we introduce the (spatially) Weighted Temporal Rich Club (WTRC) definition and demonstrate its utility in analyzing spatiotemporal interaction networks in the Experiments and Results section.
In this research, the geographic regions, including congressional districts, census tracts, and counties, are modeled as \textit{nodes}, while their spatial interactions are represented as \textit{edges}, with \textit{edge weights} representing the volume of human mobility flows between geographic regions during the specified temporal period. Building on the TRC proposed by \cite{pedreschi2022temporal}, we quantify the WTRC effect using the following steps:  
1) Identify rich nodes and calculate the average weighted connectedness of the rich club sub-network over the time duration $\Delta$ starting at $t$.  
2) Across all $t$ timestamps, determine the maximum average weighted connectedness $\max(\overline{C}_{k, \Delta})$.  
3) Normalize the maximum average weighted connectedness of the observed network by dividing it by the maximum average weighted connectedness of the null model. All the aforementioned randomization techniques, such as weight reallocation and sequence shuffling, can be used in the creation of the null model. Adapting the TRC definition from Equation~\ref{eq:normalized_M_trc}, the WTRC coefficient can be expressed as follows:

\begin{equation}
M(k, \Delta)_{norm} = \frac{\max(\overline{C}_{k, \Delta})}{\max(\overline{C}_{{k, \Delta},rand})}
\label{eq:normalized_M}
\end{equation}

\begin{figure}[]
    \centering
    \includegraphics[width=1\textwidth]{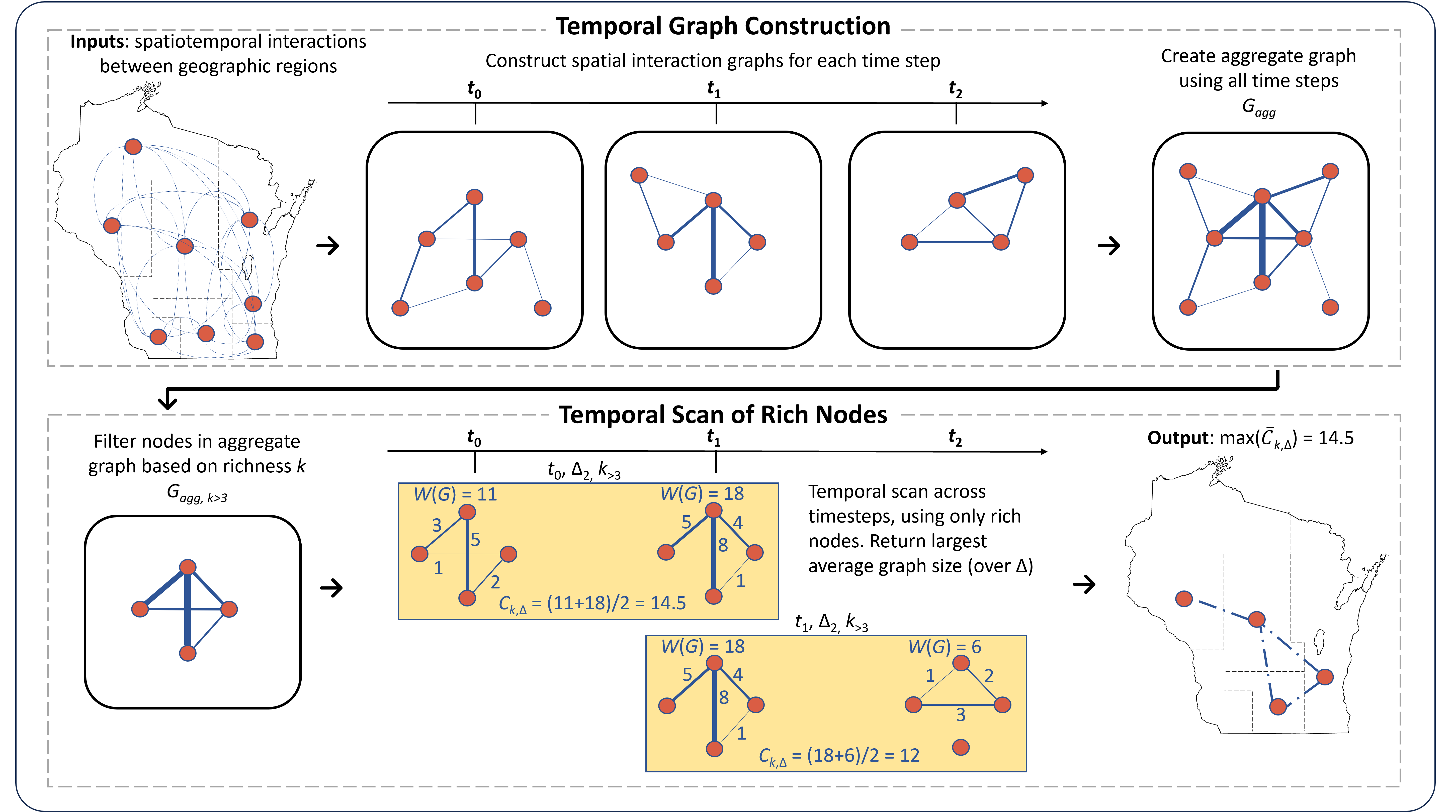}
    \caption{Steps for calculating the WTRC coefficient in a dynamic spatial interaction network. $W(G)$ represents the graph size of the rich club sub-network at a given time step. Only the numerator of Equation~\ref{eq:normalized_M} is shown, for clarity. The denominator is calculated in the same way, but on an appropriately randomized network.} 
  \label{fig:wi_wtrc_ttrc_horizontal_ave}
\end{figure}

Where $\max(\overline{C}_{{k, \Delta},rand})$ represents the average of the maximum average weighted connectedness values found across multiple null models. Although $k$ (degree) is used in Equation~\ref{eq:normalized_M} to denote the richness property, as degree is the most commonly used richness property in the literature, other suitable richness properties could be used instead.
Similarly, while we define $M(k, \Delta)_{norm}$ as the WTRC in order to maintain consistency with Equation~\ref{eq:normalized_M_trc}, we will refer to $M(k, \Delta)_{norm}$ as the WTRC coefficient for clarity.
To illustrate the specifics of the WTRC calculation, we present a toy example in Figure~\ref{fig:wi_wtrc_ttrc_horizontal_ave}, omitting the calculation of $\max(\overline{C}_{{k, \Delta},rand})$ for simplicity.

\begin{figure}[]
    \centering
    \includegraphics[width=.98\textwidth]{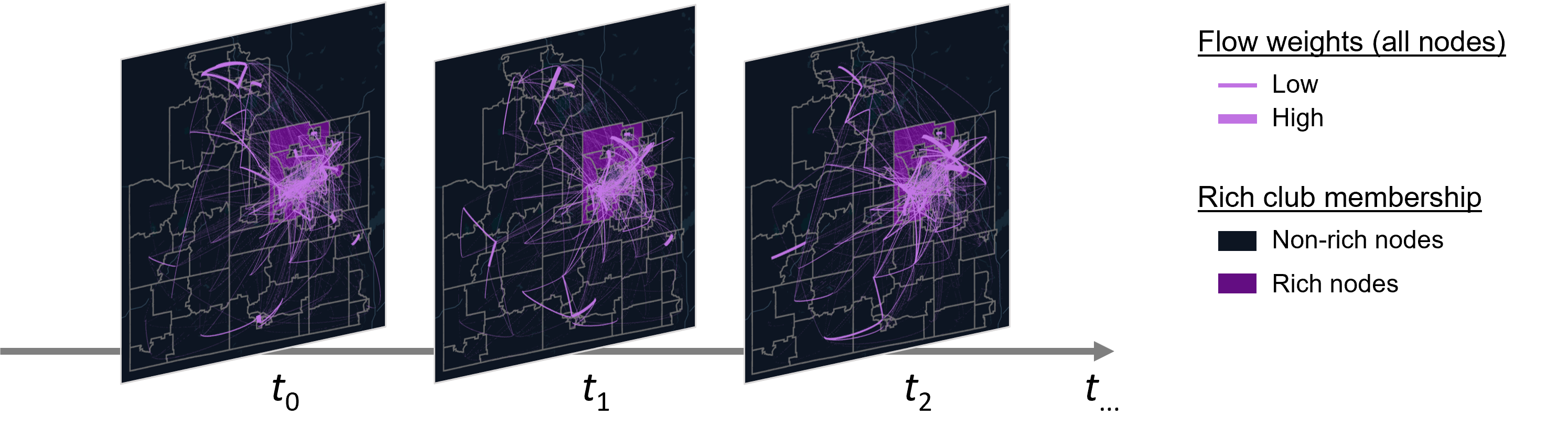}
    \caption{An example of the temporal snapshots that make up a weighted spatiotemporal network. Light purple arcs represent flows between regions, with arc thickness indicating the volume of those flows. Rich club flows are those flows that originate in a rich club region (dark purple) and end in a rich club region (dark purple).}
    \label{fig:flow_panels_dis3_cropped}
\end{figure}

While Figure~\ref{fig:wi_wtrc_ttrc_horizontal_ave} demonstrates how the WTRC is calculated at the Congressional District level for simple illustration purpose, Figure~\ref{fig:flow_panels_dis3_cropped} provides a more realistic example of the type of spatiotemporal interaction networks that the WTRC can analyze. We here describe how the WTRC would be calculated for this example network.
First, rich club members (dark purple) are identified across all temporal snapshots of the network. Then, the flows (light purple arcs) between these regions within each snapshot are used to compute the average weighted connectedness over a duration of $\Delta$ snapshots at each time step $t$, with the process being repeated for each $t$ value in a moving window fashion. Finally, the maximum of these averages is recorded.
Given the high density of flows within the rich club regions at each time step, the weighted temporal network shown in Figure~\ref{fig:flow_panels_dis3_cropped} would have a strong WTRC value.

While the proposed WTRC definition measures the size and temporal stability of a rich club in terms of weight allocation, we might also wish to know the extent to which a temporally-stable topological rich club exists in the same network, as both rich club types describe different aspects of the network. As described above, the weighted and topological rich club effects present in the same network can be distinguished by calculating each rich club type using null models that randomize either the weight allocation or the edge allocation, respectively. However, our proposed WTRC calculation is not directly comparable to the TRC proposed by \cite{pedreschi2022temporal}, as the edge stability requirement is different. Therefore, to distinguish between the topological and weighted aspects of temporal networks, we also provide a topological temporal rich club (TTRC) definition that is directly comparable with our proposed WTRC definition. Specifically, this TTRC is calculated in the same way as the WTRC, but with all edge weights in the network set to 1. This definition of the TTRC can be thought of as a more continuous measurement of the TRC introduced by \cite{pedreschi2022temporal}, in that the coefficient value of the TTRC linearly reflects the extent to which the rich club is consistently and stably interconnected, rather than only showing edges that are present at every single time step for duration $\Delta$. Finally, we note that the mixed rich club effect could be measured using the WTRC on the original network, but by randomizing both the topology and weight allocation of the network used for normalization.

With the proposed methods for measuring the weighted and topological temporal rich clubs, we now turn to experiments to demonstrate their utility in analyzing different types of time-varying spatial interaction networks.

\section{Experiments and Results}\label{sec:experiments}
\subsection{Comparing WTRC and TTRC for human mobility networks in congressional districts}\label{sec:cong_scans}
% WTRC and TTRC Congressional District scans
To demonstrate how the WTRC can reveal rich club behavior that remains hidden in purely topological or static analyses, we apply the WRC, TTRC, and WTRC to a human mobility network and compare the results. In our multiscale analysis of human mobility patterns, we begin by examining spatial interactions between census tracts in Wisconsin, with a particular focus on the networks within the state's second and third Congressional Districts for 2022. Using the WTRC and TTRC, we identify temporally stable district cores, distinguishing between topological and weighted temporal rich club effects. Although we use congressional districts to showcase our methodology and explore the relatively under-explored area of rich club analysis in redistricting, our primary goal is to demonstrate how the WTRC uncovers patterns that other methods might overlook.

Regarding the experimental settings, the richness property for both the WTRC and TTRC is defined by the temporal edge count in the time-aggregated graph. While it is common to calculate a rich club coefficient for every richness value in the sequence for the TopoRC and WRC, which have linear time and space complexity, this approach is less practical for the WTRC due to its polynomial time and space complexity.
As a result, the richness thresholds and $\Delta$ durations scanned are chosen arbitrarily, after preliminary experiments indicated that varying scan settings did not significantly alter the results. The richness thresholds are established by identifying the minimum and maximum temporal edge counts in the data, and then adding 10 evenly spaced intervals between these values, resulting in a total of 12 intervals covering the full range of temporal edge counts. The temporal edge count for a given node is defined as the sum of its instantaneous degrees over the entire temporal network (see the Discussion section for a detailed description of this richness property).
The $\Delta$ durations are determined by identifying the start and end of the temporal range (148 is used as the endpoint, even though there are 155 timestamps, to allow the largest $\Delta$ duration to span multiple timestamps). Six evenly spaced intervals are then added between these points, resulting in a total of 8 intervals covering the entire temporal range.

For randomization, we use weight decorrelation when measuring the WTRC, and edge switching when measuring the TTRC. In both cases, sequence shuffling is used to randomize the temporal dimension of the network. For the WTRC and TTRC, $\max(\overline{C}_{k, \Delta},{rand})_{mean}$ is calculated using 10 different randomized networks. Given that the rich club effect is measured across multiple $\Delta$ values and richness levels, we refer to the process as a WTRC or TTRC scan, respectively.
The implementation of the TTRC and WTRC is modified based on code from the Temporal-Rich-Club package \citep{pedreschi2022temporal}, and we use the tool \textit{kepler.gl} to visualize spatial interaction flows.

For spatial interaction data, we employ the weekly SafeGraph Neighborhood Mobility Patterns dataset, from January 7, 2019 through December 27, 2021, to estimate population-scale human mobility flows between census tracts in Wisconsin \citep{kang2020multiscale}. We construct the spatially-weighted temporal networks of human mobility flows for U.S. Congressional Districts 2 and 3 in Wisconsin.
Using the 2022 Wisconsin Congressional Districts Approved Plan\footnote{https://redistrictingdatahub.org/dataset/2022-wisconsin-congressional-districts-approved-plan/}, we perform a spatial join of the census tract geometries to the Congressional District boundaries, including census tracts where at least 60\% of their area falls within a district boundary. For each district, a spatial interaction graph is constructed for every week in the period, with \textit{nodes} representing census tracts, \textit{edges} representing spatial interactions between tracts, and \textit{edge weights} representing the volume of human mobility flows between tracts during that week (as shown in Figure \ref{fig:flow_panels_dis3_cropped}).

\begin{figure}[!tbp]
  \centering
  \subfloat[]{\includegraphics[width=0.5\textwidth]{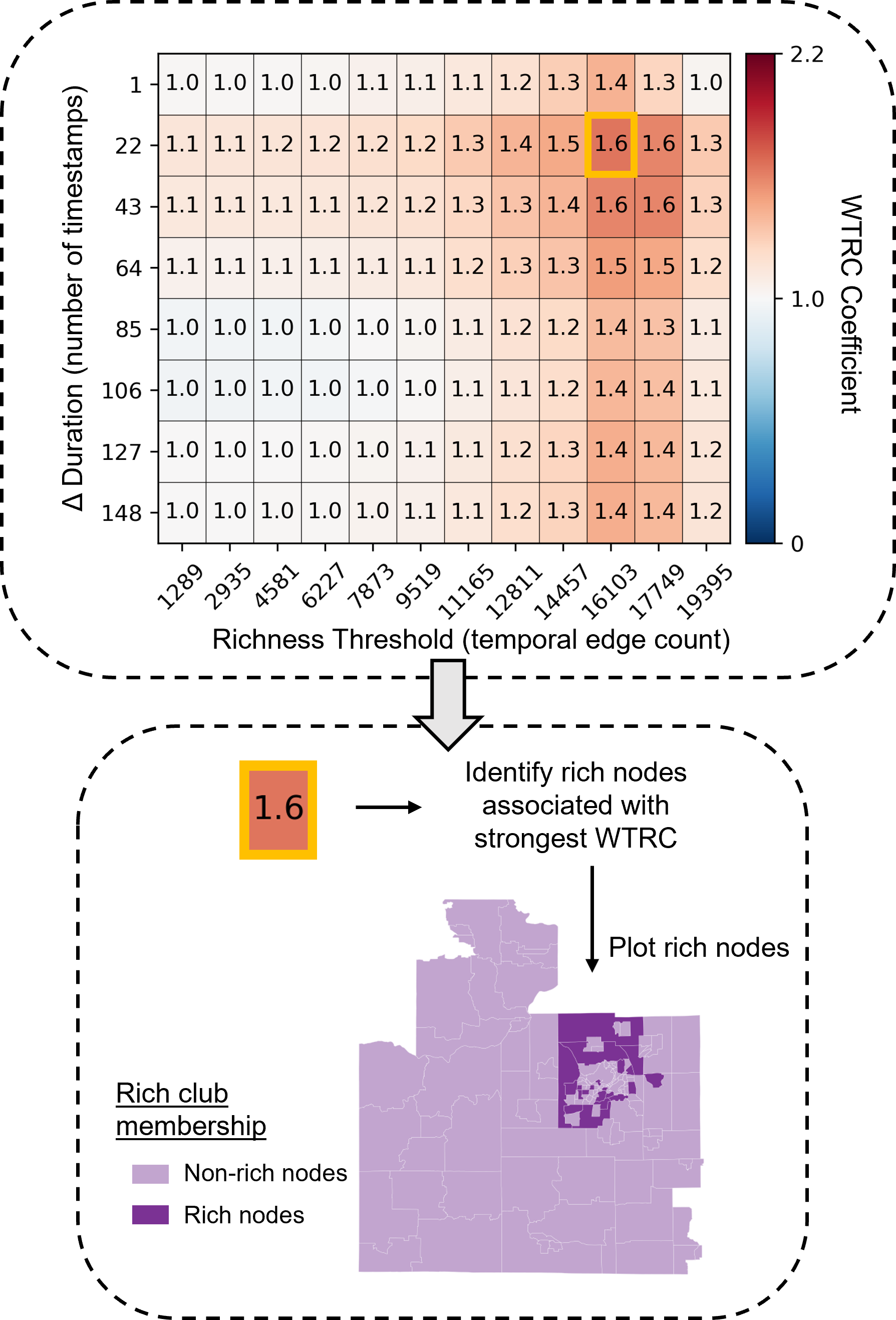}\label{fig:dis2_RCS_richNodesMap_WTRC}}
  \hfill
  \subfloat[]{\raisebox{3pt}{\includegraphics[width=0.495\textwidth]{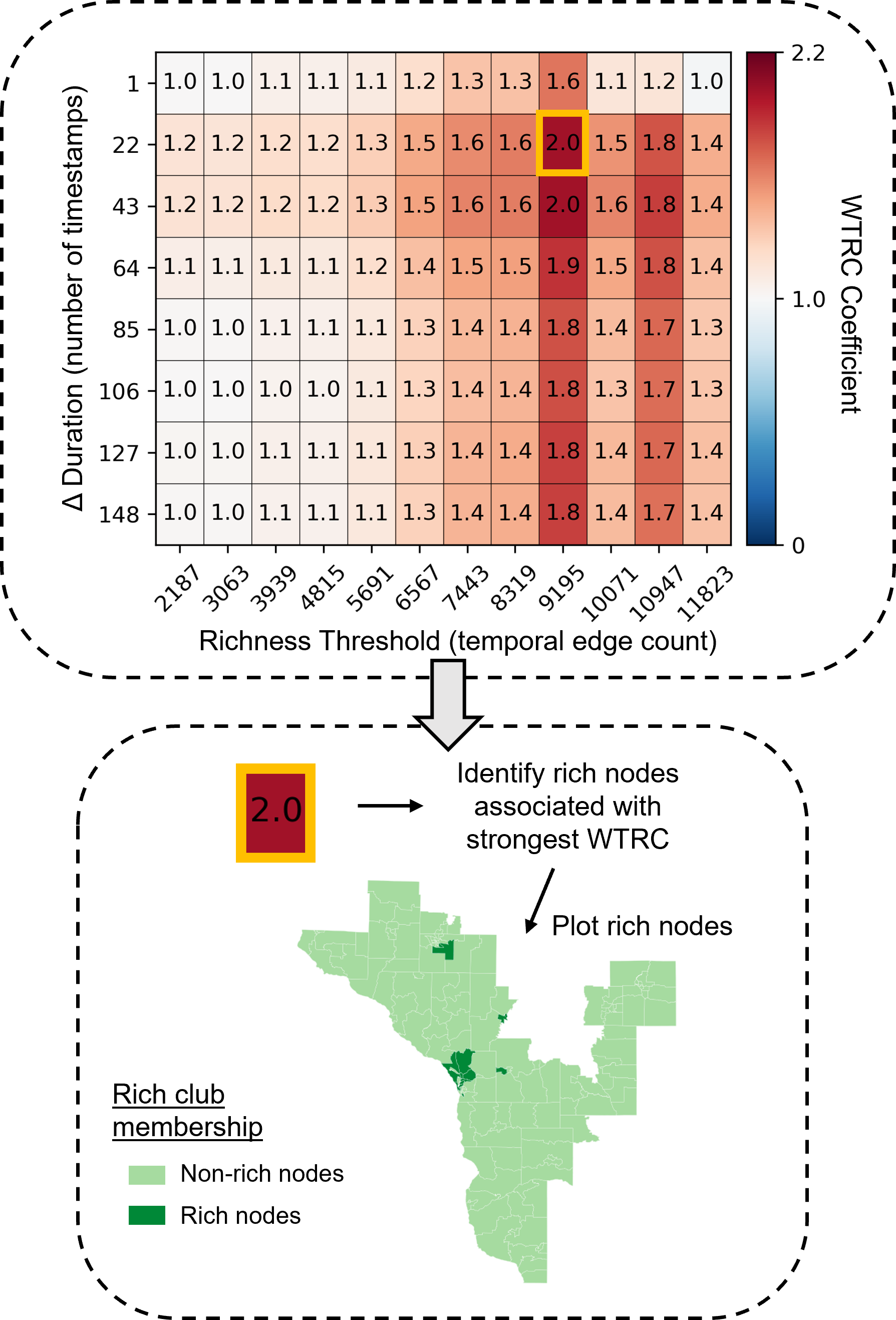}}\label{fig:dis3_RCS_richNodesMap_WTRC}}
  \caption{WTRC scan results for (a) Congressional District 2 and (b) Congressional District 3 in the State of Wisconsin. The rich-club census tracts associated with the highlighted WTRC coefficient in each Congressional District (1.6 and 2.0 respectively) are shown in dark color on the map.}
\end{figure}

\begin{figure}[!tbp]
  \centering
  \subfloat[]{\includegraphics[width=0.475\textwidth]{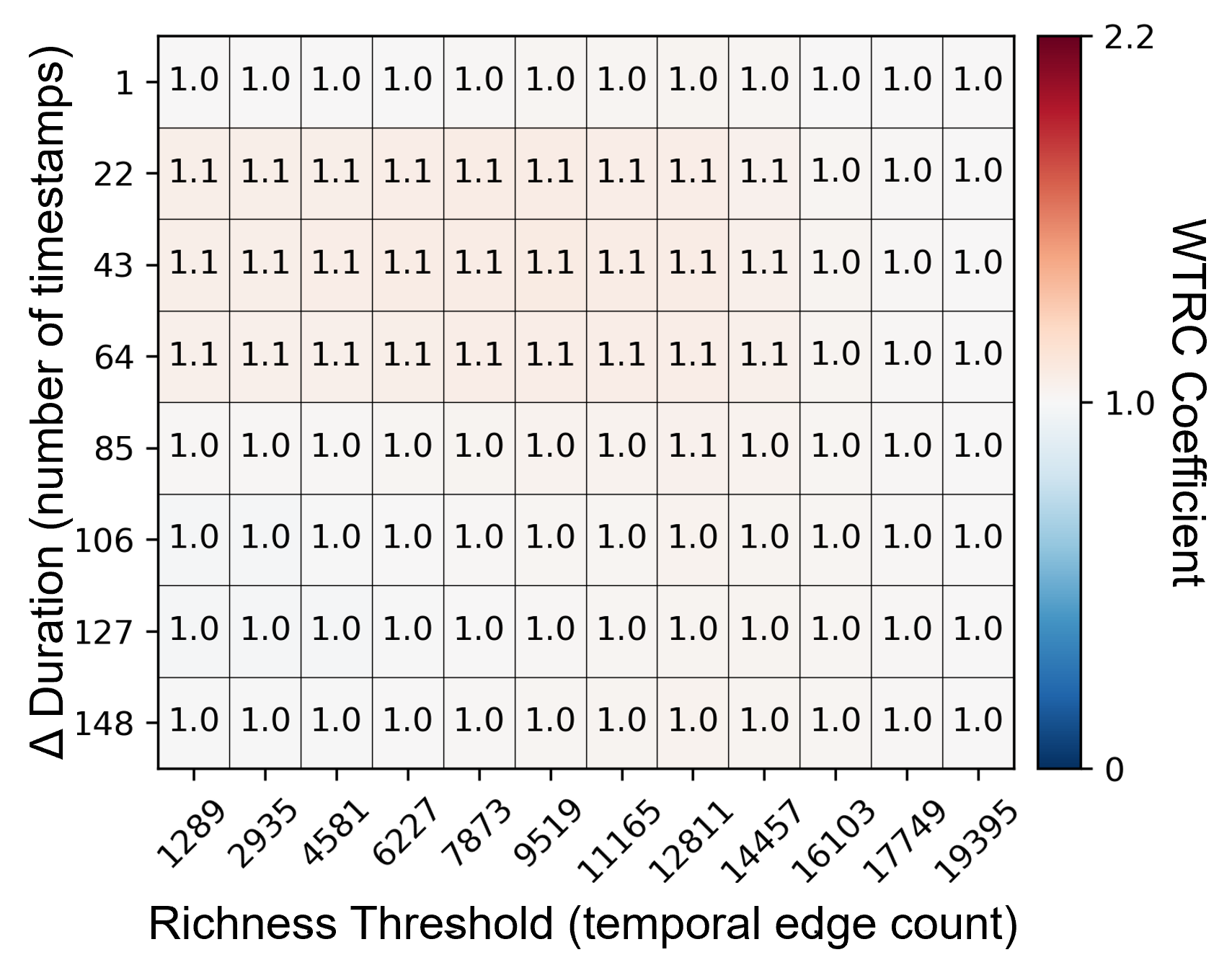}\label{fig:dis2_ttrc}}
  \hfill
  \subfloat[]{\includegraphics[width=0.48\textwidth]{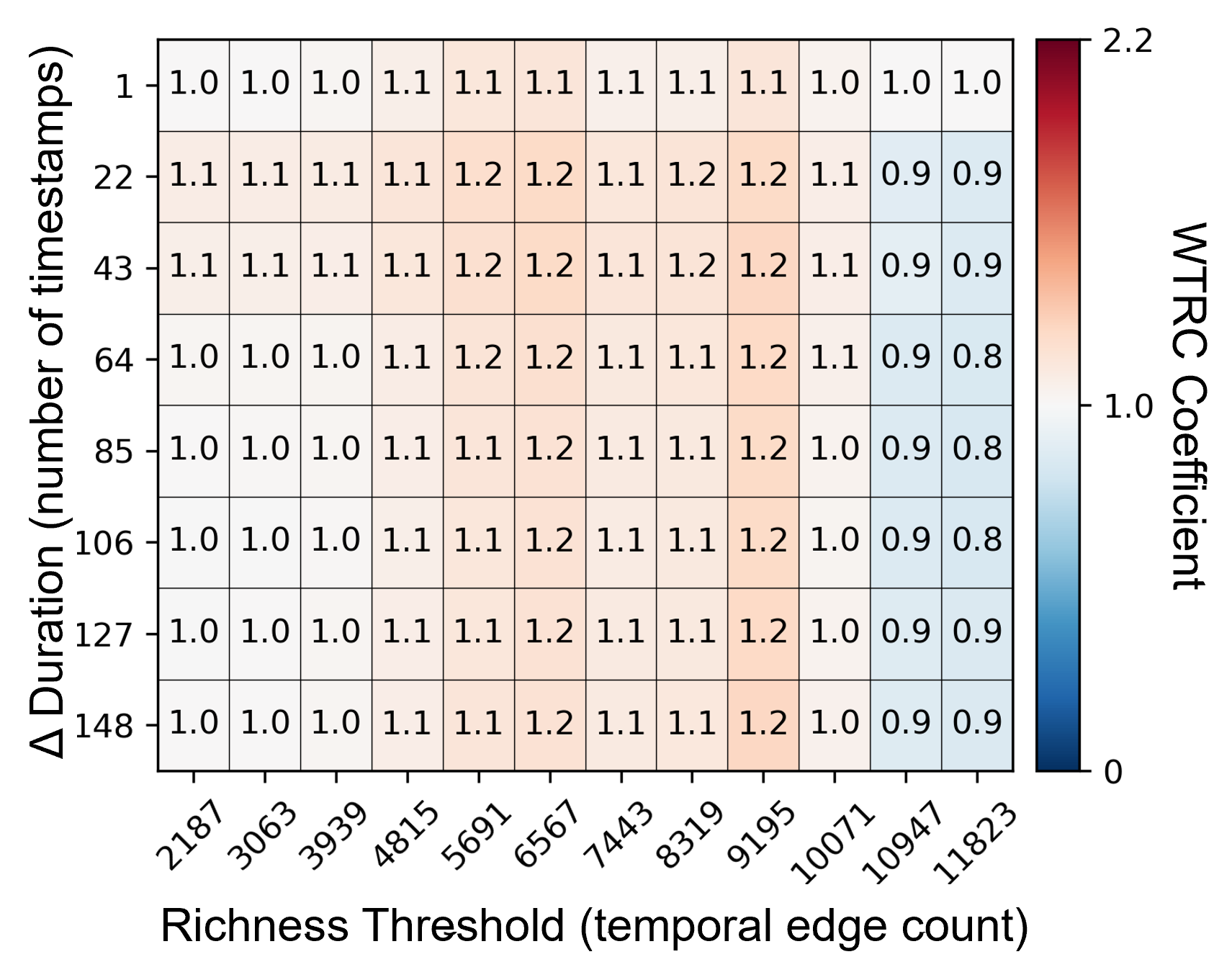}\label{fig:dis3_ttrc}}
  \caption{TTRC scan results for (a) Congressional District 2 and (b) Congressional District 3. In comparison with the weighted temporal rich clubs identified with the WTRC scans, the same Congressional districts showed little to no topological temporal rich club effect.}
\end{figure}

Using the WTRC scan, we identified strong, significant spatially-weighted temporal rich club effects in both Congressional District 2 (Figure~\ref{fig:dis2_RCS_richNodesMap_WTRC}) and Congressional District 3 (Figure~\ref{fig:dis3_RCS_richNodesMap_WTRC}), with District 3 showing a stronger WTRC effect across all richness thresholds, though its richest nodes were not as rich as those in District 2.
In terms of temporal scale, both districts exhibited the strongest WTRC effect with a $\Delta$ duration of 22 weeks (i.e., the strongest WTRC coefficient was found using a 22-week moving window to calculate the WTRC coefficient for each $t$ in $[1, T-\Delta]$). However, the strength of the WTRC coefficient remained robust even with the longest $\Delta$ duration of 148 weeks, suggesting that the spatial interactions between the richest nodes are consistent across different time scales.

Regarding richness thresholds, both districts had the strongest WTRC coefficients around the third or fourth highest richness thresholds (temporal edge counts of 16,103 for District 2 and 9,195 for District 3, respectively). 
Geographically, the rich club members in District 2 are concentrated around the city of Madison—Wisconsin's capital, while in District 3 they are based around two main cities—La Crosse and Eau Claire—with some other distant census tracts also included. 
In contrast to the WTRC scans, the TTRC scans did not identify strong rich clubs. District 2 showed almost no measurable topological temporal rich club effect, as most census tracts remained topologically connected at every time step (Figure~\ref{fig:dis2_ttrc}). District 3 exhibited only a moderate temporal rich club effect across all $\Delta$ values (Figure~\ref{fig:dis3_ttrc}). At the highest richness thresholds in District 3, the TTRC even revealed an anti-rich club effect (i.e., a TTRC coefficient $< 1$), indicating that the `richest' census tracts were less topologically connected over time than expected by chance.
In summary, the proposed method enables the detection of a weighted temporal rich club effect even in networks where no topological temporal rich club effect is present.

\subsection{Comparing the WRC and WTRC within human mobility networks in congressional districts}
While the previous section explores the differences between weighted and topological approaches to rich club analysis in a spatiotemporal network, we now shift our focus to the unique temporal insights provided by the WTRC compared to the non-temporal WRC definition. We start with this hypothesis: if the WTRC effectively captures the temporal dynamics of a network, we should observe different patterns in WTRC coefficient values across various $\Delta$ durations compared to the WRC coefficient calculated on the time-aggregated version of the same network, particularly if the data exhibits temporal variation. Put simply, using a temporal representation of the network for rich club analysis should reveal more temporal insights than a static representation.
To test this hypothesis, we plot the WTRC coefficients across the richness sequence of District 2 for selected $\Delta$ durations, alongside the static WRC coefficients calculated for the same richness sequence in the time-aggregated graph of the network. For normalization, we compare these coefficients to the average of 100 randomized versions of the time-aggregated graph, where edge weights have been decorrelated.

\begin{figure}[]
    \centering
    \includegraphics[width=.8\textwidth]{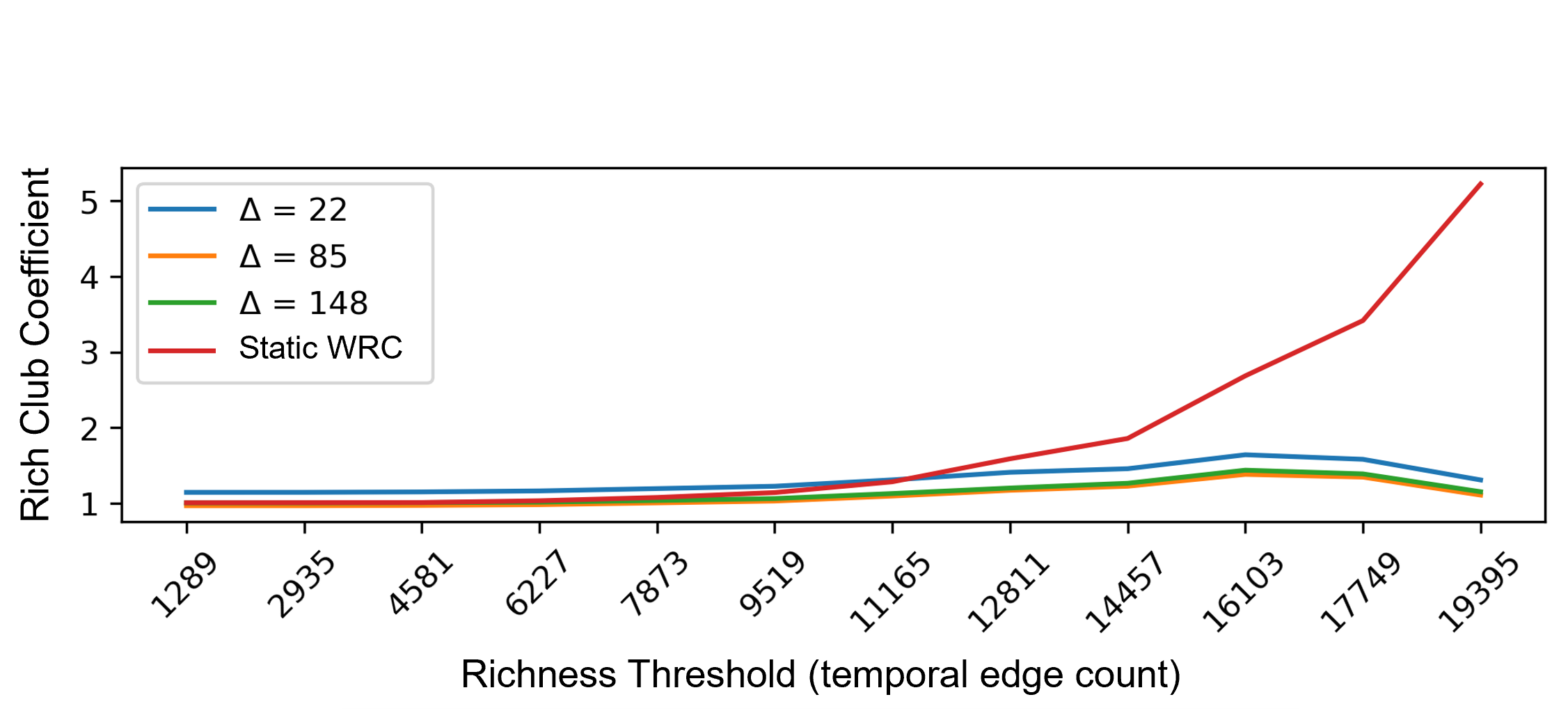}
    \caption{Static WRC values (red line) plotted against WTRC coefficient values for each richness threshold for Wisconsin Congressional District 2. Across the higher richness thresholds, the static WRC coefficient is much larger than the WTRC coefficients, showing how just using the time-aggregated graph overestimates the weighted rich club effect.}
    \label{fig:wtrc_lineplots}
\end{figure}

As seen in Figure~\ref{fig:wtrc_lineplots}, the static WRC values are similar to the WTRC values until the richness threshold of 11,165, after which they start to greatly exceed the WTRC values, ending up more than three times larger than the largest WTRC coefficient found. Furthermore, the WTRC values for $\Delta$'s 22, 85, and 148 all start to decrease after a richness threshold of 16,103 temporal edges, while the static WRC continues to increase at a greater rate. The static WRC coefficient increases continually over the richness sequence because there is indeed a WRC effect between the census tracts with the highest mobility flows, which also tend to be census tracts with the highest population (see Figure~\ref{fig:dis2_wtrc_kepler}).

However, the decreasing WTRC values across longer $\Delta$ durations show that not all the mobility flows between the high population census tracts occur over the same period, at least not to the extent that is suggested by the static WRC coefficient calculated on the time-aggregated graph. The actual WTRC effect measured is still significant, but less so than the static WRC would suggest. 
Rather than the census tracts that exceed the highest richness threshold demonstrating the strongest rich club effect, we actually see that a lower richness threshold of 16,103 temporal edges is where the strongest rich club effect is seen across all the $\Delta$ durations scanned. At that threshold, we can get even more temporal nuance by comparing the WTRC values across the three $\Delta$ durations shown in Figure~\ref{fig:wtrc_lineplots}. While the WTRC effect was weaker for both 85 and 148 weeks, it was higher for $\Delta$ = 22 (about 5 months). 

\begin{figure}[H]
    \centering
    \includegraphics[width=.8\textwidth]{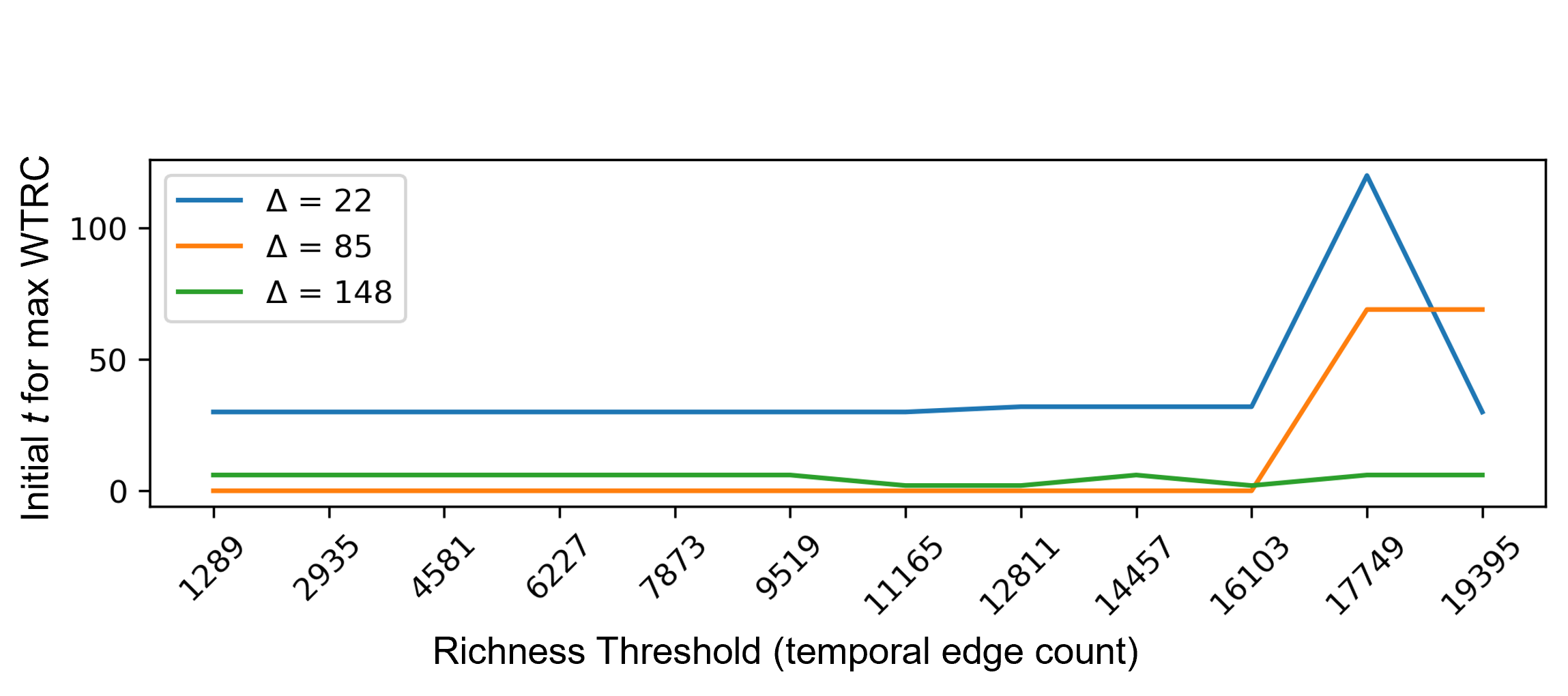}
    \caption{Initial timestamps for the maximum WTRC coefficient at each richness threshold in Wisconsin Congressional District 2. For the largest rich club coefficients, the initial timestamp varied greatly between the three different $\Delta$ durations used, indicating that rich club effect was not consistent throughout the entire period of study, but rather peaked at different times, depending on the richness threshold and $\Delta$ duration used.}
    \label{fig:wtrc_max_Ts}
\end{figure}

We can gain even further temporal insight by looking at the initial timestamps associated with the largest WTRC value found for each $\Delta$ duration and richness threshold, respectively. As seen in Figure~\ref{fig:wtrc_max_Ts}, the initial timestamp associated with each of the maximum WTRC values varied greatly between the three different $\Delta$ durations used. This variation indicates that the rich club effect was not consistent throughout the entire period of study, but rather peaked at different times, depending on the richness threshold and $\Delta$ duration used. For reference, the COVID-19 related lockdown order began in Wisconsin on March 16, 2020, which corresponds to timestamp 62 in our dataset. As seen in Figure \ref{fig:wtrc_max_Ts}, the vast majority of maximum WTRC coefficients were recorded with $\Delta$ durations that started on timestamps prior to the COVID-19 pandemic, suggesting that rich club behaviour at the census tract level was generally higher prior to the pandemic. Of note, the $\Delta$ duration of 148 is essentially the whole period, and so it is more a baseline temporal rich club effect, rather than being able to show fluctuations in the WTRC coefficient over time.  For the lower richness thresholds, the initial timestamp associated with the largest WTRC coefficients is quite stable for the $\Delta$ durations of 22 and 85. However, the maximum WTRC coefficients for $\Delta$ = 85 across the lower richness thresholds come from close to $t$=1, while for $\Delta$ = 22 these tend to come around $t$ = 40, demonstrating how the WTRC reveals temporal fluctuations that would be obscured with a static WRC analysis.

\subsection{Understanding the spatial structure of district cores using the WTRC}\label{sec:resultsampling}
\begin{figure}[H]
    \centering
    \includegraphics[width=.8\textwidth]{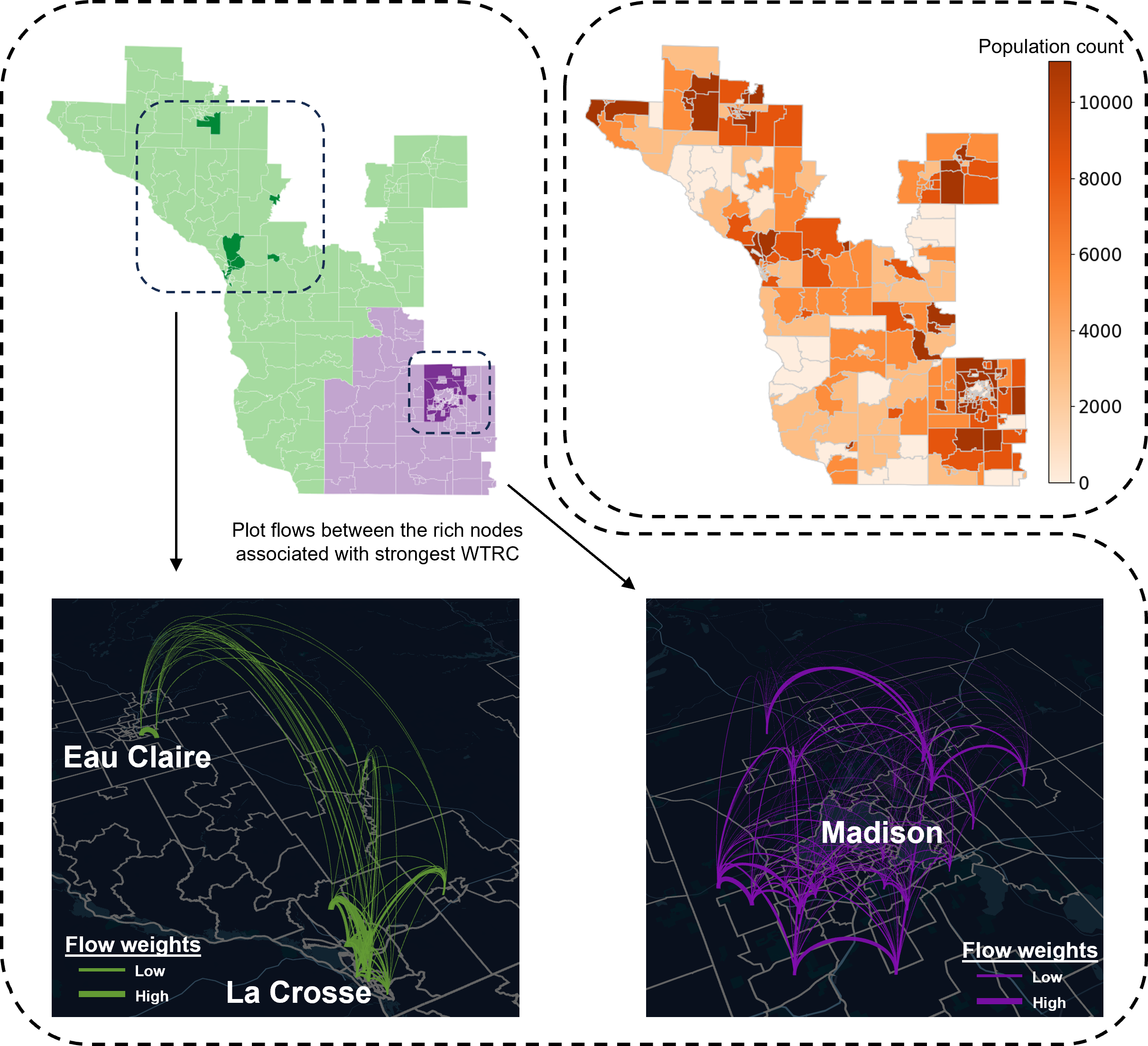}
    \caption{Rich club members and their strongest human mobility flows.
\textbf{Top Left}: The dark purple and dark green census tracts highlight the rich club members in Congressional Districts 2 and 3, respectively, associated with the strongest WTRC coefficients.
\textbf{Top Right}: Population counts for Wisconsin District 2 and 3 census tracts, with darker shades representing higher population counts. Rich club members tend to be census tracts with high populations, but this is not always the case.
\textbf{Bottom Left}: Human mobility flows between rich club census tracts in District 3, which are geographically dispersed, primarily between Eau Claire and La Crosse.
\textbf{Bottom Right}: Human mobility flows between rich club census tracts in District 2, concentrated around Madison, Wisconsin's capital.}
    \label{fig:dis2_wtrc_kepler}
\end{figure}

Turning from the temporal dimension of the WTRC to the spatial dimension of the WTRC, we now identify and analyze the rich clubs associated with strongest WTRC values found in both congressional districts. The rich club census tracts are shown in Figure~\ref{fig:dis2_wtrc_kepler}, with the darker hues in each district representing the rich club members. Comparing these two rich clubs, the most salient feature is that rich club members for District 2 are all associated with one city (Madison, the capital of Wisconsin), while the rich club members in District 3 are associated two relatively distant cities, La Crosse and Eau Claire (the two largest cities of that district). In the redistricting process, characterizing the most important substructures of the district might suggest which census units should be maintained in a district, while others could more easily be moved between districts. For example, in Figure~\ref{fig:dis2_wtrc_kepler}, it can be seen that rich club members (dark purple) are spatially clustered in the upper right portion of District 2, while census tracts that are not part of the rich club tend to be along the bottom and left side, implying that such areas could be added to another district without substantially influencing the human mobility patterns of the existing district core. For District 3, there are several large parts of the district that are not associated with the identified rich club. In particular, the large lobes on the upper right side and bottom part of the district seem as though they are quite distant from the district cores, suggesting that it might produce more coherent districts to move such regions to other districts.

Within one district, a large WTRC coefficient at a given duration $\Delta$ and richness setting means that there are strong and stable spatial interactions between the rich nodes, at least for some period of duration $\Delta$ on the interval $[1,T]$. Similarly, a larger WTRC coefficient at a longer $\Delta$ duration, compared to shorter $\Delta$ duration at the same richness level, reflects both stronger and more temporally-stable mobility flows. In the districts analyzed here, the maximum WTRC values were similar across different $\Delta$ durations at the same richness level, indicating that the spatial interaction patterns are relative stable across different times scales. Even though the initial timestamp for the WTRC of the richest nodes might vary significantly (Figure~\ref{fig:wtrc_max_Ts}), indicating when WTRC values peaked, the magnitude of the WTRC values for those nodes did not significantly change with different $\Delta$ durations.

\subsection{Temporal analysis of U.S. county mobility flows during COVID-19}
\begin{figure}[!h]
    \centering
    \includegraphics[width=1.0\textwidth]{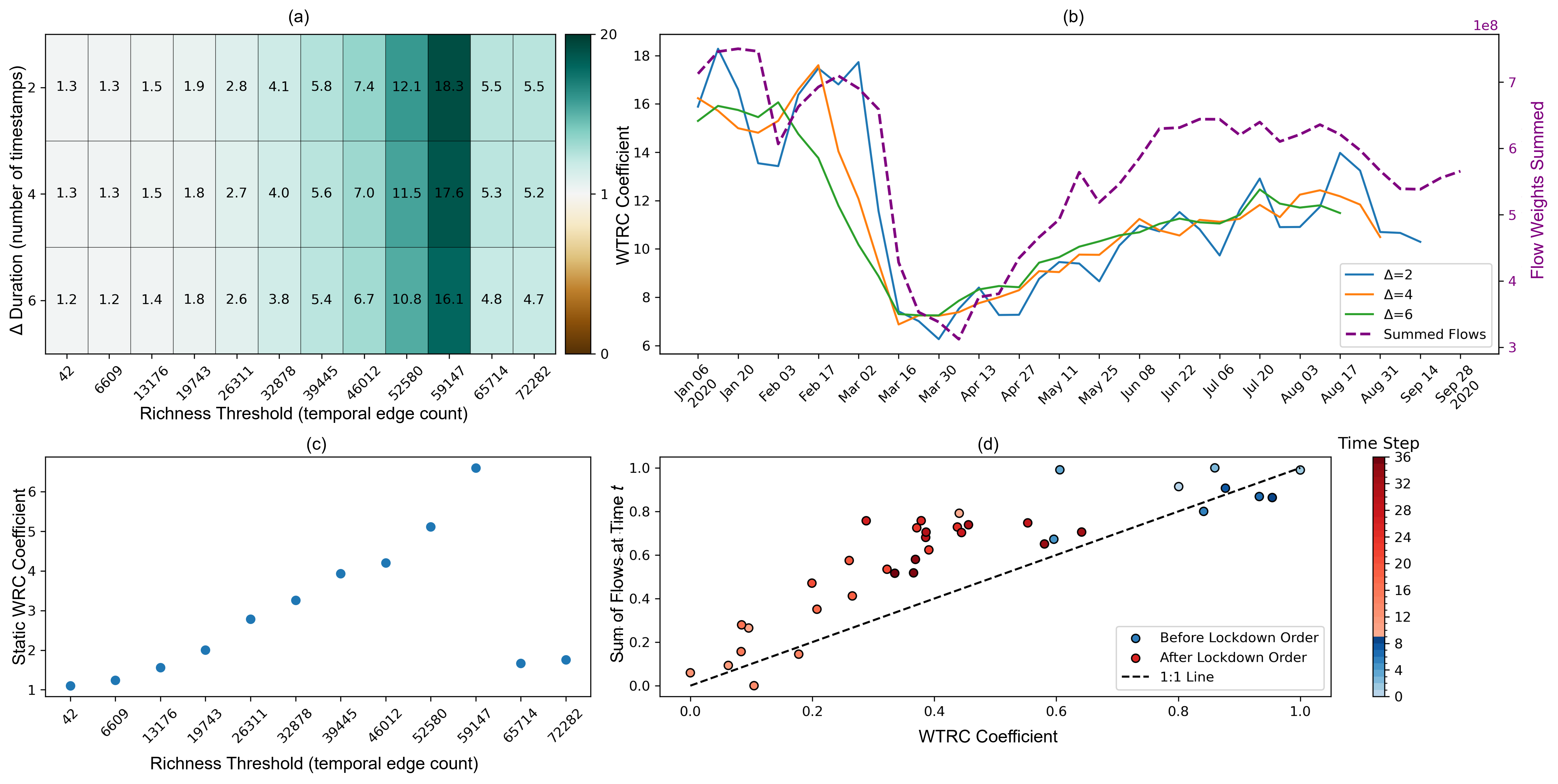}
    \caption{WTRC analysis of county-level human mobility patterns from January 2020 through September 2020: \textbf{a)} WTRC scan at three different $\Delta$ durations. \textbf{b)} The WTRC value at each timestamp $t$ across the whole time period, using the richness threshold associated with the maximum WTRC value found for each $\Delta$ duration in the scan; the sum of flows between all counties in the U.S. at each timestamp is also included for reference (second y-axis). \textbf{c)} The static WRC calculated with the time-aggregated network of U.S. counties, using the same richness thresholds as the WTRC. \textbf{d)} The WTRC values calculated at each timestamp using a $\Delta$ duration of 2 and the richness threshold of 59,147. While the sum of flows before lockdown (blue) and early in the lockdown period (light red) has a fairly 1:1 relationship with the WTRC at each timestamp, the sum of flows in the later lockdown period (dark red, corresponding to July, August, and September) is proportionally larger than the WTRC coefficient.}
    \label{fig:covid_county_panels}
\end{figure}

In the second case study of our multiscale analysis, we apply the WTRC to human mobility flows at the county level across the entire U.S., demonstrating that the WTRC is effective at capturing rich club behavior across different spatial scales.
Rather than comparing the TTRC and WTRC, as done in the previous section , we focus on using the WTRC to understand how human mobility patterns fluctuate over time, with a particular focus on changes during the outbreak of the COVID-19 pandemic~\citep{gao2020mapping}.
Using the $\Delta$ durations of 2, 4, and 6 weeks, we scan the period of January 7, 2020 through September 16, 2020, within which there exist temporal variations of human mobility patterns~\citep{kang2020multiscale}. The richness thresholds are generated in the same manner as in the first case study. In addition to presenting the maximum WTRC coefficients found over the entire study period across different $\Delta$ durations and richness thresholds, as seen in panel A of Figure \ref{fig:covid_county_panels}, we also analyze how the WTRC value varies over time. The spatiotemporal interaction network is produced in the same manner as in the first case study, using weekly SafeGraph Mobility Patterns to estimate the volume of human mobility flows, but this time using U.S. counties as the spatial unit, such that edge weights represent weekly human mobility flows and nodes represent U.S. counties. 

The WTRC scan of the dynamic spatial network of U.S. county human mobility flows revealed a strong rich club effect present across the three $\Delta$ durations used (i.e., 2, 4, and 6 weeks), all of which had a maximum value at a richness threshold of 59,147 temporal edges (Figure \ref{fig:covid_county_panels}a). Using the same richness threshold, we plot the WTRC values across the entire period for each $\Delta$ duration (Figure \ref{fig:covid_county_panels}b). As seen in the plot, the WTRC values are highest in January and February 2020, before reaching their lowest values after the lockdown orders for COVID-19 (associated with March 16, 2020). The WTRC values then slowly climb throughout the rest of the summer, but do not return to the levels seen prior to the lockdown orders. The sum of flows across all counties at each timestamp $t$ are also plotted on the second y-axis of Figure~\ref{fig:covid_county_panels}b, and they follow roughly the same pattern as the WTRC values. Notably, however, the sum of flows increases more than the WTRC values in the summer of 2020. 

To investigate the increase in the sum of flows relative to the WTRC, we plot the WTRC values for the richness threshold of 59,147 and $\Delta$ duration of 2 weeks against the sum of flows at the same time step (Figure \ref{fig:covid_county_panels}d). Furthermore, the timestamp of each point is color-coded so that points before the COVID lockdown orders are blue and points after the lockdown orders are red. Earlier points in each period are shown with a lighter hue. The first aspect of interest is that the pre-COVID points (blue) have the highest flow sums and highest WTRC values, and the ratio between the values is close to 1:1. Early on after the lockdown orders (light red), the sum of flows and the WTRC still have a fairly 1:1 relationship, but their absolute values are much lower, corresponding to the drop in human movements associated with the lockdown orders. After March 16, 2020, however, the flows and WTRC values increase gradually, peaking in August 2020. Interestingly, the sum of flows in the later lockdown period (dark red, corresponding to July, August, and September) is proportionally larger than the WTRC coefficient. This suggests that people tended to add trips to counties that were \textit{not} a part of the rich club during summer 2020, relative to the pre-COVID baseline. Finally, we note that none of these temporal dynamics are captured by the static WRC calculated on the time-aggregated network (Figure \ref{fig:covid_county_panels}c), which does not and cannot show temporal fluctuations in the rich club effect, like those that occurred around the COVID-19 lockdown orders.

\begin{figure}[H]
    \centering
    \includegraphics[width=1.0\textwidth]{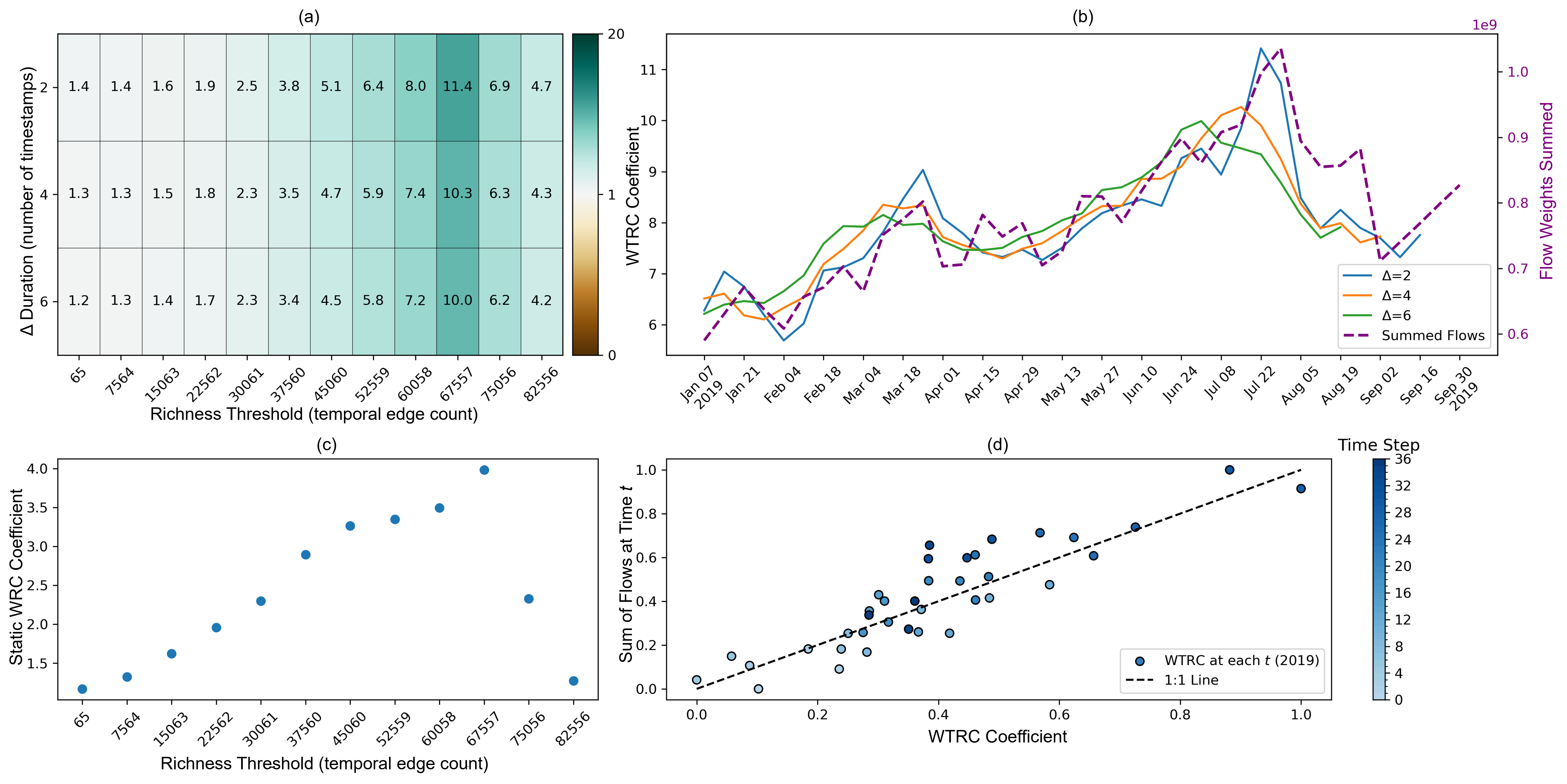}
    \caption{WTRC analysis of county-level human mobility patterns over 2019: \textbf{a)} WTRC scan at three different $\Delta$ durations. \textbf{b)} The WTRC value at each timestamp $t$ across the whole time period, using the richness threshold associated with the maximum WTRC value found for each $\Delta$ duration in the scan; the sum of flows between all counties in the U.S. at each timestamp is also included for reference (second y-axis). \textbf{c)} The static WRC calculated with the time-aggregated network of U.S. counties, using the same richness thresholds as the WTRC. \textbf{d)} The WTRC values calculated at each timestamp using a $\Delta$ duration of 2 and the richness threshold of 67,557, plotted against the sum of flows. } 
    \label{fig:us_counties_2019}
\end{figure}

To ensure that the human mobility changes observed around COVID-19 are not merely normal seasonal fluctuations, we also perform the same analysis described above but using the human mobility flows in 2019 over the same months of the year. As seen in Figure \ref{fig:us_counties_2019}a, the maximum WTRC values found in 2019 are substantially lower than those found in 2020 (the maximum WTRC values in 2020 come from the pre-lockdown period; Figure \ref{fig:covid_county_panels}b), even while the maximum sum of flows for 2019 (in August) is significantly higher than the maximum found in 2020 (in January). As seen in Figures \ref{fig:covid_county_panels}b and d, the relationship between WTRC values and the sum of flows at each $t$ is relatively linear throughout 2019.  

To quantitatively compare the mobility patterns in 2019, pre-lockdown 2020, and post-lockdown 2020, we further build three simple linear regression models, all using standardized values to allow for comparison. Specifically, these models are: 1) a linear model that predicts 2019 WTRC values ($Y$) with 2019 weekly human mobility flows ($X$); 2) a linear model that predicts pre-lockdown 2020 WTRC values with 2020 pre-lockdown weekly human mobility flows; and 3) a linear model that predicts post-lockdown (i.e., March 16, 2020 and on) WTRC values with post-lockdown weekly human mobility flows. As seen in Table \ref{tab:regression_output}, the 2019 flows have a strongly linear relationship, 0.78, with the WTRC values of that period. The pre-lockdown period in 2020 has a weakly linear relationship, but that may be due partially to data sparsity. For the post-lockdown flows, there is a strong, linear relationship of 0.87 with WTRC values--a significant increase over the relationship between flows and the WTRC seen in 2019 and in pre-lockdown 2020. The coefficient of 0.87, versus 0.78, represents an approximately 12\% increase in the WTRC relative to the sum of flows between the two periods, meaning that the weighted temporal rich club effect in the time-varying spatial network increased significantly after the lockdown orders, even while the overall mobility flows decreased.

\begin{table}[H]
\centering
\begin{tabular}{lcc}
\hline
Time Period & Slope (Correlation Coefficient) & Intercept \\
\hline
2019 Flows & 0.78 & 0.07 \\
Pre-lockdown 2020 Flows & 0.35 & 0.00 \\
Post-lockdown 2020 Flows & 0.87 & 0.00 \\
\hline
\end{tabular}
\caption{Linear regression model parameters for models built for each of the three periods, with the sum of flows predicting the WTRC value for each $t$ as the dataset. All datasets were normalized using min-max scaling.}
\label{tab:regression_output}
\end{table}

\begin{figure}[H]
    \centering
    \includegraphics[width=.85\textwidth]{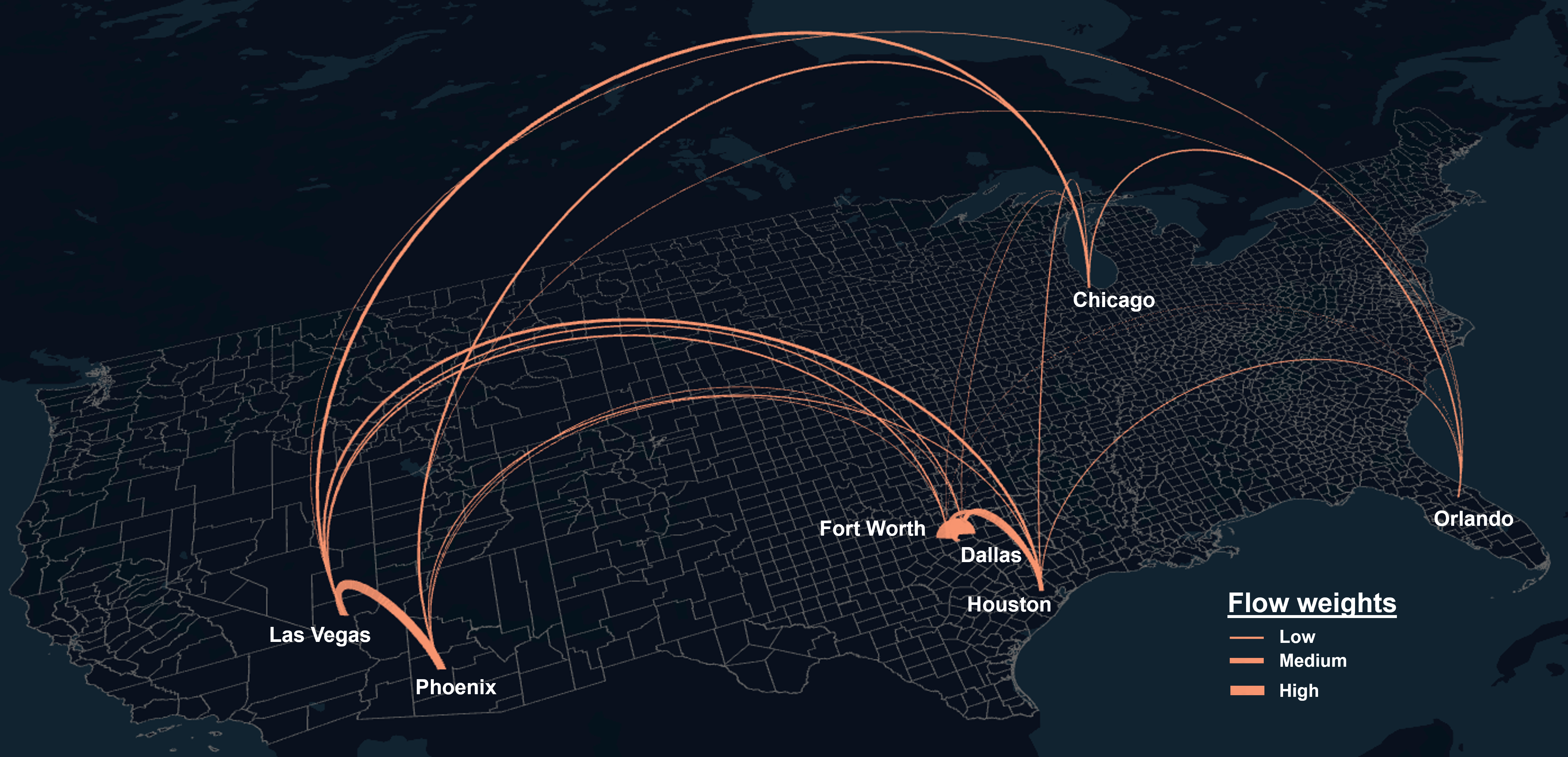}
    \caption{Flows between members of the rich club counties (using a richness threshold of 59,147 temporal edges) in March 2020. The name of the city to which the rich club county pertains is labeled on the map. Except in the case of Dallas and Fort Worth, which are also geographically-adjacent cities, the rich club counties visualized here correspond to the major airports of each city, and thus the rich club visualized here is largely comprised of air traffic-related human mobility flows at the national scale.} 
    \label{fig:us_counties_covid}
\end{figure}

Turning now to the spatial structure of the strongest rich club identified in year 2020, we also visualize the rich club of counties above the 59,147 temporal edge count richness threshold (Figure \ref{fig:us_counties_covid}). As seen with the city labels, the counties correspond to major U.S. cities, and in fact each of counties at the 59,147 temporal edge count richness threshold are home to their city's major airport. Therefore, the WTRC scan at the 59,147 temporal edge count richness threshold and county-level geographic scale is mostly capturing human mobility flows related to air traffic between major transportation hubs. The one exception to this might be in the case of Dallas-Fort Worth, as these are two densely populated cities that are adjacent to each other. Indeed, the flows between the central counties of these two cities at any given timestamp are, on average, an order of magnitude larger than the flows between any of the other rich club members, which is true both before and after the lockdown order in March 2020. The dichotomy between the relatively short distance between Dallas and Fort Worth and the much greater distances between other cities (e.g., Orlando to Chicago) highlights the varying `cost' of travel across different spatial scales \citep{barthelemy2011spatial}. This cost encompasses factors such as travel time, expenses. etc. For example, car travel is well-suited for connecting geographically adjacent rich club members, offering lower costs for short distances, but becomes increasingly expensive and less practical for regional connections. Conversely, air travel significantly reduces costs and travel times for longer, regional distances but proves less efficient and more costly when applied to local travel needs.

\section{Discussion}\label{sec:discussion}

\subsection*{Alternative richness definitions}\label{supplementary}
In two of the existing studies using the topological TRC, the degree in the time-aggregate graph has been used as the richness property \citep{pedreschi2022temporal,li2023temporal}, where the degree $k$ of node $i$ in the time-aggregate graph $G$ is the number of unique nodes with which $i$ has shared an edge at least once on the interval $[1,T]$.
\cite{niu2023abnormal} used only one $k$ value for their TRC scan of the human brain. 
In a static, weighted context, \cite{alstott2014unifying} used the degree sequence as well. 

\begin{figure}[H]
	\centering
	\includegraphics[width=.8\textwidth]{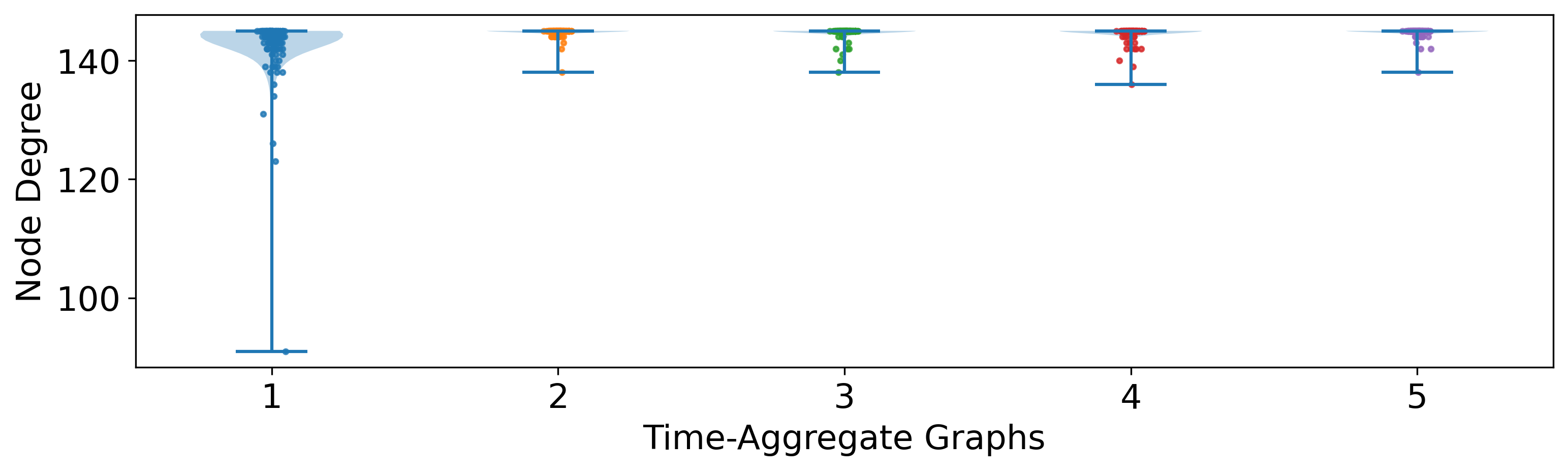}
	\caption{In the time-aggregate graph ($T=155$)  census tracts ($n=146)$ in the second Congressional District of Wisconsin, there are very few unique node degree values. If node degree was used as the richness sequence, as done in other works, the would be little range or diversity in richness values. This is true of the original temporal network (label 1), and even more so for the networks randomized with edge switching (labels 2-5). } 
	\label{fig:aggK_dist_cts_dis2}
\end{figure}

In highly connected temporal networks, however, using the degree in the aggregate graph may not provide a diverse enough richness sequence to allow for distinction between various richness levels. Consider as an example the node degrees in the time-aggregate graph for human mobility flows within Wisconsin census tracts in Congressional District 2. As seen in Figure~\ref{fig:aggK_dist_cts_dis2}, most of the 146 census tracts in the original network (label 1) have close to the maximum degree possible. In the versions of the network randomized with edge switching (labels 2-5), the range of unique node degree values in the time-aggregate graph narrows even further. Furthermore, the edge switching randomization algorithm also changes the number of rich nodes (defined by degree) at each threshold, violating the requirement the null model have the same maximal possible connectedness within the rich club \citep{alstott2014unifying}.

\begin{figure}[H]
	\centering
	\includegraphics[width=.8\textwidth]{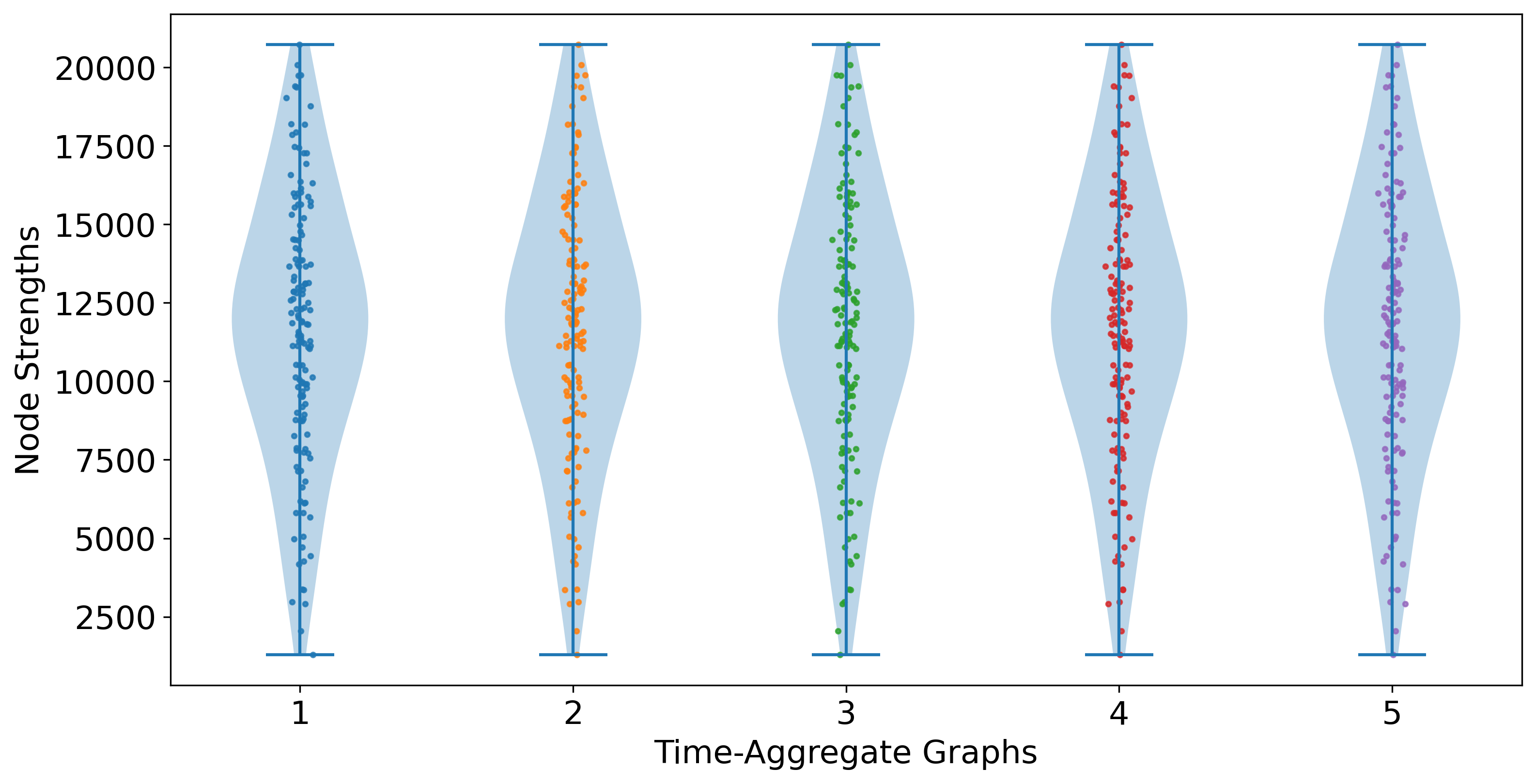}
	\caption{Temporal edge counts (node strengths) for census tracts in the second Congressional District of Wisconsin. Label 1 pertains to the original network, and labels 2-5 pertain to networks randomized with edge switching. Notably, the sequence of temporal edge counts does not change with randomization.}
	\label{fig:ttrc_temp_edge_count}
\end{figure}

Possible richness sequence alternatives to degree are network properties that are preserved after both weight allocation and topological randomization, allowing for direct comparison of the WTRC and TTRC effects at the same $\Delta$ durations and specific richness thresholds.
One such alternative for the richness sequence is the \textit{total temporal edge count} for each node, i.e., the sum of instantaneous degrees across all timestamps for a given node. Equivalently, this is the node strength of the time-aggregate graph in unweighted temporal networks. As seen in Figure~\ref{fig:ttrc_temp_edge_count}, there is a much broader distribution in these values compared to the static degree distribution in the aggregate graph seen in Figure~\ref{fig:aggK_dist_cts_dis2}, while the unique values are preserved both in the original network (label 1) and in networks randomized with edge switching (labels 2-11). Weight decorrelation randomization, similarly, does not change the temporal edge count for each node in the time aggregate graph, as the topology remains unchanged in the weight decorrelation method. 

\subsection*{Broader Implications}\label{broader}
Having demonstrated the utility of the WTRC for spatiotemporal network analysis, we now discuss various aspects of this work.
Regarding the WTRC effect across different spatial scales, the WTRC coefficients calculated using county-level human mobility flows were significantly larger than those derived from census tract-level flows, even though similar time periods were analyzed. As shown in Figure \ref{fig:us_counties_covid}, the pronounced rich club effect at the county level appears to be driven by airports located within the rich club counties. In contrast, mobility within congressional districts is likely dominated by car traffic, given the minimal regional air traffic within Wisconsin. This multiscale analysis demonstrates that the WTRC is robust across various spatial scales and highlights how different spatial processes can be revealed at each scale.

Although we have used the temporal edge count as the richness property, other richness properties could be utilized. If distinguishing between topological and weighted temporal rich club effects is desirable, the richness property chosen should remain consistent after both edge placement and weight allocation randomization to allow for meaningful comparisons between the two effects for a given set of rich nodes. Additionally, while sequence shuffling is employed to disrupt all temporal patterns in the null models, preserving certain temporal patterns may be beneficial in specific domains—an aspect that will be explored in future work. Furthermore, parameters such as the number of $\Delta$ durations and richness thresholds scanned could significantly impact the results of a WTRC analysis. To ensure robustness, it is crucial to test various scan settings, as was done in the preliminary work for this research.

Building on the discussion of spatially-explicit networks from the Introduction section, we provide further justification for why many of the rich club networks studied in the literature can be classified as spatial networks.
As \cite{janowicz2020geoai} described, an invariance test can be used to determine if a model is spatially-explicit: if the results of the model vary under relocation of the studied phenomena, then the model is considered to be spatially-explicit. Similarly to how a permutation test can be used with Moran’s I to determine the statistical significance of the observed spatial autocorrelation, the null models used for rich club normalization are used to evaluate whether or not the observed allocation of weights and edges is more clustered than expected by chance (e.g., if the weights and/or edges were shuffled). The fact that rich club values can change with this type of normalization reveals that the WTRC is in fact spatially-explicit, even if rich club research in complex networks does not typically come from a spatial perspective.

\section{Conclusion and Future Work}\label{sec:conclusion}
In this research, we have presented a novel method called the (spatially) weighted temporal rich club (WTRC), to effectively quantify the temporal rich club phenomenon in undirected, weighted spatiotemporal interaction networks. The new approach is especially useful for analyzing dynamic spatial interaction networks in which the topological connections are temporally stable, but varying in their spatial interaction weights.
For analyzing temporal networks, the WTRC can provide insights about the temporal scale and variability of rich clubs, which would otherwise be obscured in a static rich club analysis of the same network represented in a static form.
Even in spatial networks without strong temporal patterns, the WTRC coefficient can be used to quantify the precise strength of the WRC effect across different richness thresholds, yielding insight into the extent to which rich club members direct spatial interaction flows to each other. This novel approach has significant implications for GIScience and beyond, offering a powerful tool to analyze complex dynamic spatial interaction networks at different spatial scales across various domains, such as epidemiology, transportation, and redistricting.  

\section*{Data and Code Availability}
The data and codes that support the findings of this study are openly available at the following GitHub repository: \url{https://github.com/GeoDS/WTRC/} . 

\section*{Declaration of interest statement}
The authors declare no competing interests. 

\section*{Acknowledgments}
The authors are grateful to Nicola Pedreschi and Daniel Szabo for their insightful feedback and valuable discussions on the temporal rich club and rich club normalization methods. We acknowledge the data support from SafeGraph and the funding support by the National Science Foundation (No. 2327797) and the University of Wisconsin 2020 WARF Discovery Initiative project: Multidisciplinary Approach for Redistricting Knowledge. Any opinions, findings, and conclusions or recommendations expressed in this material are those of the author(s) and do not necessarily reflect the views of the funders.

\section*{Author Biographies}
\noindent JACOB KRUSE is a PhD candidate in the Department of Geography at the University of Wisconsin-Madison, WI 53706. E-mail: jikruse@wisc.edu. His research interests include redistricting, public health, geospatial data science and GeoAI.

\noindent SONG GAO is an Associate Professor of Geographic Information Science at the Department of Geography, University of Wisconsin-Madison, WI 53706. E-mail: song.gao@wisc.edu. His research interests include geospatial data science, GeoAI, social sensing, spatial networks and human mobility.

\noindent YUHAN JI is a PhD candidate in the Department of Geography at the University of Wisconsin-Madison, WI 53706. E-mail: yuhan.ji@wisc.edu. Her research interests include transportation, human mobility, geospatial data science and GeoAI.

\noindent KEITH LEVIN is an Assistant Professor in the Department of Statistics at the University of Wisconsin-Madison, WI 53706. E-mail: kdlevin@wisc.edu. His research interests include network analysis, dimension reduction, concentration inequalities and clustering problems.

\noindent QUNYING HUANG is a Professor at the Department of Geography, University of Wisconsin-Madison, WI 53706. E-mail: qhuang46@wisc.edu. Her research interests include geospatial data science, remote sensing, spatial computing and natural hazards.

\noindent KENNETH MAYER is a Professor at the Department of Political Science, University of Wisconsin-Madison, WI 53706. E-mail: krmayer@wisc.edu. His research interests include redistricting and American Politics.

\bibliographystyle{apalike}
\bibliography{references}

\end{document}